\documentclass[twocolumn,English,aps,pra,10pt,superscriptaddress,floatfix]{revtex4-2} 

\usepackage{textcomp}
\usepackage{graphicx}
\usepackage{dcolumn}
\usepackage{bm}
\usepackage{bbm}

\usepackage{times}
\usepackage{graphicx}
\usepackage{amssymb}
\usepackage{amssymb,amsfonts,amsmath,amsbsy,t1enc,latexsym}
\usepackage{float}
\usepackage[colorlinks=true,citecolor=blue,linkcolor=magenta]{hyperref}
\usepackage[markup=blue, authormarkupposition=left]{changes}
\usepackage{url}
\usepackage[mode=text]{siunitx}
\usepackage{soul}
\usepackage{changes}
\usepackage{cancel}
\usepackage{times}
\usepackage{chemformula,xspace} 
\usepackage{braket}
\usepackage[capitalise]{cleveref} 

\begin{document}


\title{Steady-state dynamics of quantum frequency combs in microring resonators}

\author{Patrick Tritschler}
\email{patrick.tritschler@de.bosch.com}
\affiliation{Robert Bosch GmbH, Robert-Bosch-Campus 1, Renningen, 71272, Germany}
\affiliation{Institute for Micro Integration (IFM), University of Stuttgart, Allmandring 9b, Stuttgart, 70569, Germany}

\author{Torsten Ohms}%
\affiliation{Bosch Sensortec GmbH, Gerhard-Kindler Stra{\ss}e 9, Reutlingen, 72770, Germany}

\author{Andr\'{e} Zimmermann}%
\affiliation{Institute for Micro Integration (IFM), University of Stuttgart, Allmandring 9b, Stuttgart, 70569, Germany}
\affiliation{Hahn-Schickard, Allmandring 9b, Stuttgart, 70569, Germany}

\author{Peter Degenfeld-Schonburg}%
\affiliation{Robert Bosch GmbH, Robert-Bosch-Campus 1, Renningen, 71272, Germany}

\date{\today}

\begin{abstract}

Optical frequency combs are utilized in a wide range of optical applications, including atomic clocks, interferometers, and various sensing technologies. They are often generated via four-wave mixing in chip-integrated microring resonators, a method that requires low optical input power due to the high-quality factor of the resonator, making it highly efficient. While the classical properties of optical frequency combs are well established, this work investigates the quantum-mechanical characteristics of the individual comb modes. We derive closed-form analytical expressions describing the squeezing, second-order correlation and joint spectral intensity between the generated signal and idler modes. This comprehensive theoretical framework enables an intuitive understanding and optimization of the quantum features across the comb, revealing conditions for substantial squeezing and entanglement relevant for quantum information processing. Our findings highlight the profound impact of design and dispersion on these quantum properties and offer a foundational tool for chip-integrated quantum applications, including quantum sensing, computing and communication.

\end{abstract}

\maketitle

\section{Introduction}

\begin{figure}
	\centering\includegraphics[width=8.6cm]{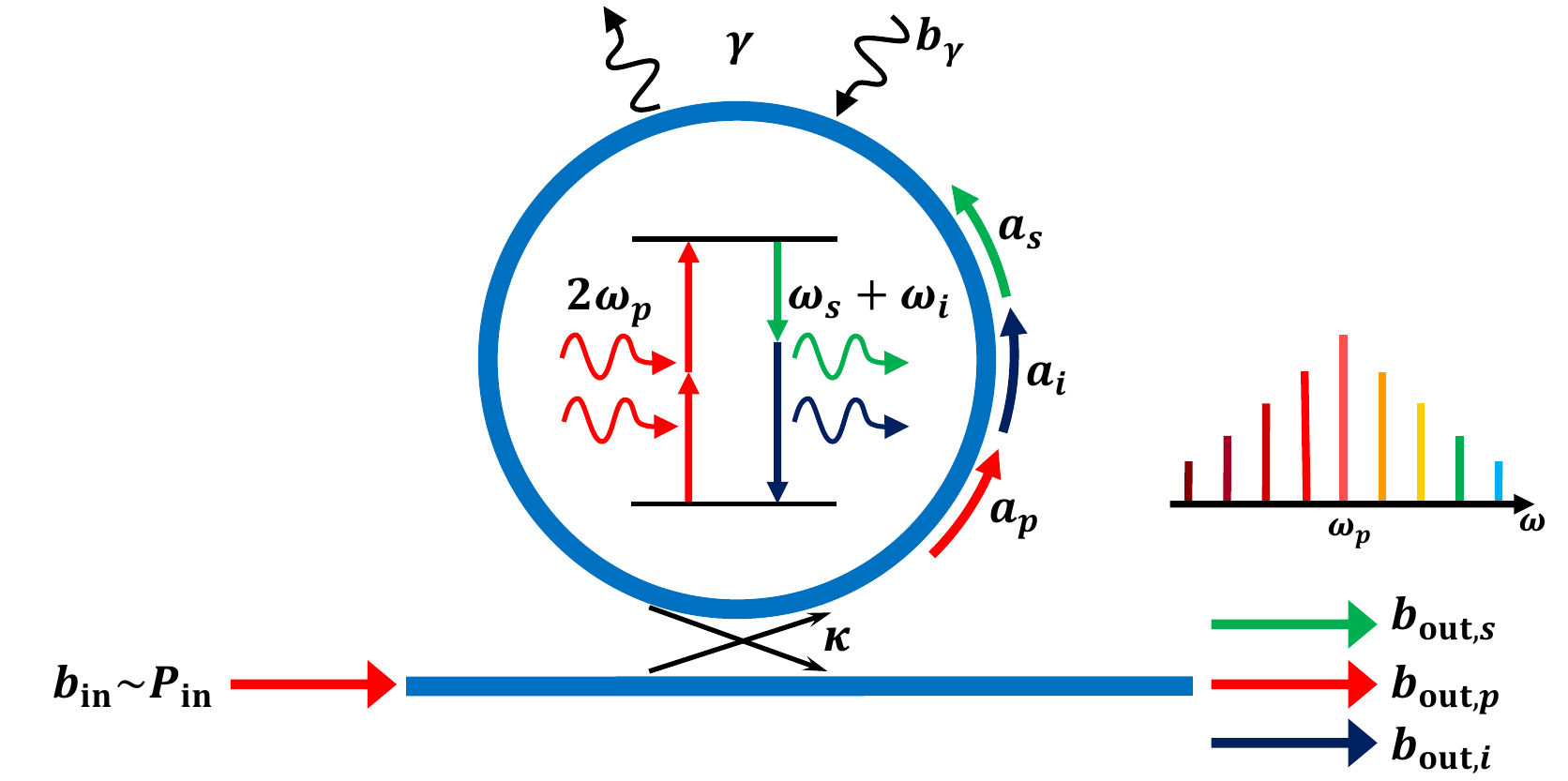}
	\caption{Schematic setup of a microring resonator which is used to generate a frequency comb via FWM. The input mode $b_\mathrm{in}$ with the optical power $P_\mathrm{in}$ couples via the coupling rate $\kappa$ into the resonator and forms the resonator mode $a_p$ that receives losses through the loss rate $\gamma$ and $\kappa$. Via FWM two modes $a_p$ are absorbed and a signal and idler mode pair $a_s$ and $a_i$ are emitted that form the outcoupled frequency comb consisting of the modes $b_\mathrm{out,p}$, $b_\mathrm{out,s}$ and $b_\mathrm{out,i}$.}
	\label{fig:ringresonator}
\end{figure}

Optical frequency combs consist of multiple, evenly spaced frequencies within a single wave packet. These distinct characteristics are highly valuable for a wide range of applications, including atomic clocks \cite{Papp:14, Zhang2015, Fortier_2024}, optical spectroscopy \cite{PhysRevLett.107.233002, Picque2019}, lidar \cite{Kuse2019}, and other sensing technologies \cite{Horiuchi2024}, mostly by exploiting their spectral properties. \\
In modern applications, optical frequency combs are often generated in microring resonators via four-wave mixing (FWM), where two pump photons are absorbed to generate a signal and an idler mode \cite{Del’Haye2007, Kippenberg2011-uh, Herr2012}. In its simplest form, a microring resonator consists of a ring-shaped optical waveguide coupled to a straight waveguide, as depicted in Figure \ref{fig:ringresonator}. With a proper design, a small input power $P_\mathrm{in}$ in the straight waveguide can lead to a high optical power within the resonator \cite{micro_ringresonators}. An optical frequency comb is generated inside the ring resonator when the pump power $P_\mathrm{in}$ exceeds a certain optical threshold power $P_\mathrm{th}$ and the energy and momentum conservation conditions for the FWM process are fulfilled \cite{Herr2012}. A crucial benefit of using microring resonators for optical frequency-comb generation is their compact form factor combined with a low threshold power $P_\mathrm{in}$, as detailed in \cite{Chang2020}. Depending on the ring resonator's design, a frequency comb with specific spectral characteristics can be measured at the output of the straight waveguide \cite{FujiiTanabe+2020+1087+1104}. The classical behavior of optical frequency combs is well understood and comprehensive overviews of their generation and applications can be found in \cite{Herr2012, Fortier2019}.\\
Beyond their classical features, recent works also explored the quantum behavior of optical frequency combs generated via FWM. It has been shown that the modes within the frequency comb can exhibit squeezing and entanglement \cite{Liu:16, Yang2021, Guidry2022, Strekalov_2016, PhysRevLett.114.050501, PhysRevLett.108.083601, PhysRevLett.101.130501, Caspani2017, Lu2019, Cheng2023, tritschler2025chipintegratedsqueezedlight}. These quantum features are of particular interest for applications in quantum sensing, computing, and communication. For instance, squeezing can be exploited to achieve noise reduction in sensing applications \cite{PhysRevA.93.053802, Michael2019, tritschler2024optical, 2025Herman}, famously demonstrated in gravitational wave detection \cite{Goda2008, Vahlbruch_2010, Jia_2024}. Similarly, entanglement is crucial for fault-tolerant photonic quantum computing \cite{PhysRevLett.101.130501, Bourassa_2021, PhysRevA.107.052414, Arrazola2021} and for realizing quantum communication protocols \cite{Lu2019}. Typically, only the quantum features between one signal and one idler mode are utilized. However, if an optical frequency comb is generated with quantum features across multiple modes, it is often termed a quantum frequency comb (QFC) \cite{Liu:16, Reimer2016, Yang2021, Guidry2022, Strekalov_2016, PhysRevLett.114.050501, PhysRevLett.108.083601}. \\
Despite the established concept of QFCs, simple closed-form equations describing the quantum properties of each mode in a chip-integrated microring resonator, particularly their dependence on geometry and a description are missing. Therefore, this work aims to derive theoretical equations that describe the dynamics of a quantum frequency comb generated in a microring resonator below threshold. We start first by modeling the classical frequency-comb generation in section \ref{cha:classicalDynamics} and then proceed to the quantum dynamics in section \ref{cha:quantumDynamics}, starting with linearized equations to describe the output modes. This enables us to derive equations for the squeezing, second-order correlation and joint spectral intensity (JSI) of the generated modes. Subsequently, our results are discussed in Section \ref{sec:discussion}, followed by a summary in Section \ref{sec:summary-and-outlook}. \\
We demonstrate that the frequency detuning of individual modes and thus the dispersion of the ring resonator, is critical for both the classical behavior of the frequency combs and their quantum features. Most interestingly, while only a few modes with negligible frequency detuning exhibit a high photon number, a significant number of modes still show substantial quantum behavior. This indicates that significant quantum features of the optical modes can be achieved in both the normal and anomalous dispersion regimes of the ring resonator. Crucially, depending on the specific design, either the mode or particle entanglement of particular optical modes can be enhanced or reduced. These findings lay the foundation for developing miniaturized, chip-integrated photonic solutions for quantum light sources, which are essential for quantum-enhanced systems and emerging quantum technology applications.

\section{Frequency comb generation in microring resonators}
\label{frequencyCombGeneration}

In this section we start by introducing the microring resonator system. We then proceed with a classical description of frequency-comb generation in section \ref{cha:classicalDynamics}, with the primary goal to understand the detuning of a resonator mode as a function of the optical input power, the ring geometry, and the material parameters. Subsequently, the quantum dynamics are introduced in section \ref{cha:quantumDynamics}, where equations describing the squeezing spectrum, the second-order correlation function, and the JSI are derived. \\
Using the system structure of the microring resonator illustrated in Fig. \ref{fig:ringresonator}, we present our model for frequency-comb generation. A laser with the pump mode $b_\mathrm{in}$, amplitude $b_\mathrm{in}=\sqrt{P_\mathrm{in} / \hbar\omega_p}$ and angular frequency $\omega_p$ is launched into the straight waveguide. This pump light couples into the ring resonator via the coupling rate $\kappa$, forming the resonator pump mode $a_p$, which is affected by losses through the loss rate $\gamma$ and by out coupling via $\kappa$. A high-power resonator mode $a_p$ is achieved when $\omega_p$ matches the ring resonator's resonance frequency $\omega_{R,T_0}=2\pi c m / n_\mathrm{eff} L_\mathrm{eff}$ at a reference temperature $T_0$, where $c$ is the speed of light in vacuum, $m$ is the mode number, $n_\mathrm{eff}$ is the effective refractive index and $L_\mathrm{eff}$ is the effective ring length \cite{micro_ringresonators}. A high-power resonator mode facilitates the generation of nonlinear effects, such as frequency-comb formation. To describe these effects and the dynamics of the optical modes within the resonator, we utilize the well-known Hamiltonian of the FWM process, given by
\begin{equation}\label{equ:fourWaveMixing_H}
\begin{aligned}
	H_\mathrm{FWM} = &\hbar \left( \omega_{p,T_0} a_p^\dagger a_p + \omega_{s,T_0} a_s^\dagger a_s + \omega_{i,T_0} a_i^\dagger a_i\right) \\
	& - i\hbar g_\mathrm{opt}\left( a_p a_p a_s^\dagger a_i^\dagger - a_p^\dagger a_p^\dagger a_s a_i \right)\\
	& - \frac{\hbar g_\mathrm{opt}}{2} \left( a_p^\dagger a_p^\dagger a_p a_p + a_s^\dagger a_s^\dagger a_s a_s + a_i^\dagger a_i^\dagger a_i a_i \right) \\
	& - 2\hbar g_\mathrm{opt} \left( a_s^\dagger a_s a_p^\dagger a_p + a_i^\dagger a_i a_p^\dagger a_p + a_s^\dagger a_s a_i^\dagger a_i \right).
\end{aligned}
\end{equation}
The first line represents the resonance frequencies of the relevant pump ($a_p$), signal ($a_s$), and idler ($a_i$) modes. The second line describes the FWM process, the third line the self-phase modulation (SPM) process and the last line the cross-phase modulation (XPM) \cite{PhysRevA.92.033840}. SPM describes the nonlinear interaction within a single mode, while XPM describes the nonlinear intensity-intesity interaction between different modes. Both effects cause a frequency detuning of the cold cavity resonance frequencies $\omega_{p,T_0}$, $\omega_{s,T_0}$, and $\omega_{i,T_0}$, as will be shown in more detail in section \ref{cha:classicalDynamics}. The intensity of the nonlinear behavior, such as a FWM process, depends on the optical nonlinearity of the waveguide, quantified by $g_\mathrm{opt} = \hbar \omega_p^2 v_g^2 n_2 / c A_\mathrm{eff} L_\mathrm{eff}$, where $v_g$ is the group velocity, $A_\mathrm{eff}$ is the effective mode area and $n_2$ is the nonlinear refractive index \cite{PhysRevA.92.033840, tritschler2024optical}.\\
While some other works include sum terms for multiple signal and idler modes within the Hamiltonian of equation \ref{equ:fourWaveMixing_H}, in this work we implicitly include all signal and idler modes within $a_s$ and $a_i$ and differentiate between individual modes later using their respective detunings. Thus, we consider a model where the pump mode couples to each signa-idler mode pair separately but the different signal-idler mode pairs do not interact with each other. Due to the low photon numbers below threshold and the rather weak optical nonlinearities $g_\mathrm{opt}$, this rotating-wave-like approximation is well valid for all practical purposes. The Hamiltonian of equation \ref{equ:fourWaveMixing_H} forms the foundation for describing the classical dynamics of the frequency comb in section \ref{cha:classicalDynamics}, followed by the quantum dynamics in section \ref{cha:quantumDynamics}.

\subsection{Classical dynamics}
\label{cha:classicalDynamics}

\begin{figure*}
	\includegraphics[width=8.5cm]{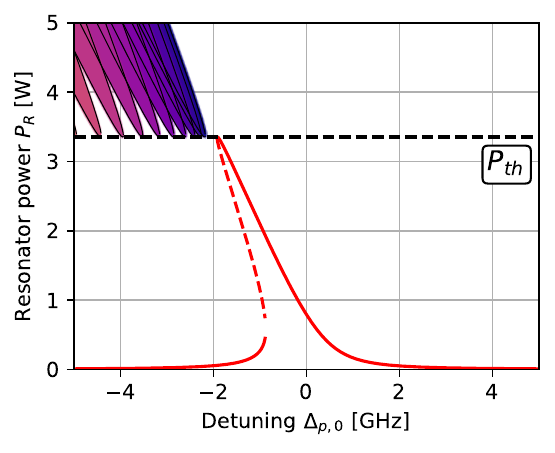}
	\includegraphics[width=8.5cm]{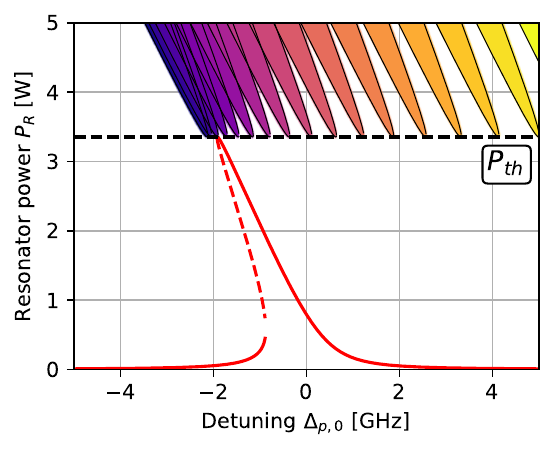}
	\caption{\label{fig:dispersion} Energy conservation of the FWM process in microring resonators in dependence of the detuning of the bare detuning of the pump mode $\Delta_{p,0} = \omega_{p} - \omega_{p,T_0}$ for various mode numbers $\mu$ and with $\kappa=300$ MHz, $\gamma=200$ MHz, $g_\mathrm{th}=10$ Hz, $g_\mathrm{opt}=1.5$ Hz and $\omega_p=1550$ nm. The red line indicates the optical power of the pump mode $P_p$, while the colored areas correspond to the required threshold power of a mode $\mu$, starting with $\mu=0$ for the blue area. The black dashed line indicates the minimum required threshold power $P_\mathrm{th}$. \textbf{Left:} Normal dispersion ring resonator with $D_2=-60$ MHz. \textbf{Right:} Anomalous dispersion ring resonator with $D_2=60$ MHz.}
\end{figure*}

For the investigation of frequency-comb generation, we begin with a classical description. While this procedure is well known and the complex mode interactions of classical frequency combs above threshold are typically modeled using the Lugiato-Lefever equation \cite{Coen:13, Lugiato_2018}, our derivation aims to analyze the detuning of individual resonator modes in greater detail. As we will demonstrate in section \ref{cha:quantumDynamics}, this detailed analysis is crucial for accurately describing the quantum behavior of the generated signal and idler modes.\\
We start our analysis by splitting each mode $a_j$ into a classical amplitude $\alpha_j$ and a fluctuating part $\delta a_j$ such that $a_j = \alpha_j + \delta a_j$. In this section our focus is solely on the classical component $\alpha_j$. Therefore, we utilize the Hamiltonian from equation \ref{equ:fourWaveMixing_H} and state the well-known classical equations below threshold \cite{Gardiner85, Collet84, tritschler2024optical}, including damping $\Gamma = \kappa + \gamma$ introduced by the coupling rate $\kappa$ and the optical loss rate $\gamma$, as
\begin{equation}
\begin{aligned}
	\frac{d \alpha_s}{d t} = i \Delta_s\alpha_s - \frac{\Gamma}{2} \alpha_s - i g_\mathrm{opt} \alpha_p^2 \alpha_i^*, \label{equ:eqm_signal}
\end{aligned} 
\end{equation}
\begin{equation}
\begin{aligned}
\frac{d \alpha_i}{d t} = i \Delta_i\alpha_i - \frac{\Gamma}{2} \alpha_i - i g_\mathrm{opt} \alpha_p^2 \alpha_s^*, \label{equ:eqm_idler}
\end{aligned} 
\end{equation}
\begin{equation}
\begin{aligned}
\frac{d \alpha_p}{d t} = i \Delta_p\alpha_p - \frac{\Gamma}{2} \alpha_p + \sqrt{\kappa}b_\mathrm{in}, \label{equ:eqm_pump}
\end{aligned} 
\end{equation}
including the introduction of the frequency detunings with
\begin{equation}
\begin{aligned}
\Delta_s& = \omega_s - \omega_{s,T_0} + g_{\mathrm{opt}} |\alpha_s|^2 + 2 g_{\mathrm{opt}} \left(|\alpha_p|^2 + |\alpha_i|^2 \right) \\
& + g_{\mathrm{th}} \left(|\alpha_{p}|^2 + |\alpha_{s}|^2 + |\alpha_{i}|^2 \right), \label{equ:signal_Detuning}
\end{aligned} 
\end{equation}
\begin{equation}
\begin{aligned}
\Delta_i &= \omega_i - \omega_{i,T_0} + g_{\mathrm{opt}} |\alpha_i|^2 + 2 g_{\mathrm{opt}} \left(|\alpha_s|^2 + |\alpha_p|^2 \right) \\
&+ g_{\mathrm{th}} \left(|\alpha_{p}|^2 + |\alpha_{s}|^2 + |\alpha_{i}|^2 \right), \label{equ:idler_Detuning}
\end{aligned} 
\end{equation}
\begin{equation}
\begin{aligned}
\Delta_p &= \omega_p - \omega_{p,T_0} + g_{\mathrm{opt}} |\alpha_p|^2 + 2 g_{\mathrm{opt}} \left(|\alpha_s|^2 + |\alpha_i|^2 \right) \\
&+ g_{\mathrm{th}} \left(|\alpha_{p}|^2 + |\alpha_{s}|^2 + |\alpha_{i}|^2 \right). \label{equ:pump_Detuning}
\end{aligned} 
\end{equation}
While the pump mode in the resonator $\alpha_p$ is increased by the input mode $b_\mathrm{in}$, the signal and idler modes $\alpha_s$ and $\alpha_i$ increase in dependence of $\alpha_p$ and the optical nonlinearity $g_\mathrm{opt}$. However, each mode is attenuated due to the losses $\Gamma=\gamma+\kappa$ and due to the detunings $\Delta_j$ that are described with equations \ref{equ:signal_Detuning}-\ref{equ:pump_Detuning}. Each detuning equation includes the SPM and XPM effects, which depend on the amplitudes of all interacting modes. In the following, we focus on an operation below threshold where $\alpha_p \gg \alpha_s, \alpha_i$ and thus we neglect the SPM and XPM influences caused by the signal and idler modes. It is important to note that for complex multimode dynamics the SPM and XPM influences needs to be included for each mode individually, which however requires a numerical investigation and is beyond the scope of this work.\\
Following \cite{PhysRevLett.134.123802}, we have further included thermal influences within the frequency detuning terms using the thermal nonlinearity $g_\mathrm{th} = \hbar\omega_p^2 n_\mathrm{eff}\gamma_{\mathrm{abs}} a_\mathrm{th} / 2 k L_\mathrm{eff}$, where $\gamma_{\mathrm{abs}}$ is the absorption loss rate, $a_\mathrm{th}$ is the temperature coefficient and $k$ is the thermal conductivity. Consequently, the optical and thermal nonlinearities contribute to optical and thermal SPM and XPM effects, respectively.\\
To analyze the dynamics of the classical FWM process, we combine the equations of motion for the signal modes in the matrix form
\begin{equation}
\begin{aligned}
\frac{d}{dt}\left(\begin{matrix} 
\alpha_{s}  \\
\alpha_{i}^* \\
\end{matrix} \right) = \left(\begin{matrix} 
i \Delta_{s} - \frac{\Gamma}{2} & -ig_\mathrm{opt}\alpha_{p}^2  \\
ig_\mathrm{opt}{\alpha_{p}^*}^2  &  -i \Delta_{i} - \frac{\Gamma}{2} \\
\end{matrix} \right) \left(\begin{matrix} 
\alpha_{s}  \\
\alpha_{i}^* \\
\end{matrix} \right) = \mathbf{L}\cdot \left(\begin{matrix} 
\alpha_{s}  \\
\alpha_{i}^* \\
\end{matrix} \right).
\end{aligned}
\end{equation} \label{equ:classical_eqm}
We utilize the nontrivial solution arising from the singularity of the matrix $\mathbf{L}$, which marks the FWM threshold. By setting $\det\left(\mathbf{L}\right)=0$ and solving for $\alpha_{p}$, we determine the necessary pump mode threshold power to
\begin{equation}\label{equ:threshold_simp}
	\alpha_\mathrm{th}^2 = \frac{\sqrt{\Delta_i \Delta_s + \Gamma^2/4}}{g_\mathrm{opt}}.
\end{equation}
This equation is not the final solution for the threshold power, since it still depends on each mode via the detuning terms $\Delta_i$ and $\Delta_s$. However, in the case of perfect resonance these detunings in equation \ref{equ:threshold_simp} vanish an the minimum required threshold for the FWM process can be determined from the steady-state solution of equation \ref{equ:eqm_pump} and the definition of $b_\mathrm{in}$, resulting in
\begin{equation}\label{equ:threshold_p}
P_\mathrm{th} = \frac{\Gamma^3\hbar\omega_{p}}{8g_\mathrm{opt}\kappa}.
\end{equation} 
To receive a more proper solution including the detunings, it is required to use the energy conservation of the FWM process $2\omega_p = \omega_i + \omega_s$ and to express the angular frequency of the signal mode in dependency of the resonance frequencies with
\begin{equation}\label{equ:signalWavelength}
\omega_s = \omega_{p} - \frac{\omega_{i,T_0}}{2} + \frac{\omega_{s,T_0}}{2}.
\end{equation}
This result can be combined with the Taylor series of the dispersion in microring resonators $\omega_\mu = \omega_0 + D_1 \mu + 1/2 D_2 \mu^2$ from \cite{FujiiTanabe+2020+1087+1104} to derive the frequency detuning for a resonator mode only in dependence of the pump mode. Therefore, we assume a symmetric behavior between the signal and idler mode, which leads to $\Delta_i = \Delta_s = \Delta_{\mu}$ and introduce the detuning of mode $\mu$ with
\begin{equation}\label{equ:signalDetuning}
\Delta_{\mu} = \Delta_{p,0} - \frac{1}{2} D_2 \mu^2+ 2g_\mathrm{opt}|\alpha_{p}|^2+ g_\mathrm{th}|\alpha_{p}|^2
\end{equation} 
and the bare frequency detuning of the pump mode $\Delta_{p,0} = \omega_{p} - \omega_{p,T_0}$. Combining equations \ref{equ:threshold_simp} and \ref{equ:signalDetuning} leads to the solution of the required pump amplitude in the ringresonator in dependence of the mode number and frequency detuning with
\begin{equation} \label{equ:critical_power}
\small
\begin{aligned}
|\alpha_\mathrm{th}|^2 &= \frac{\left( D_2\mu^2 - 2\Delta_{p,0} \right) \left( 2g_\mathrm{opt} + g_\mathrm{th} \right)}{6g_\mathrm{opt}^2 + 8 g_\mathrm{opt} g_\mathrm{th} + 2g_\mathrm{th}^2} \\
&\pm \frac{\sqrt{ g_\mathrm{opt} \left( 2\Delta_{p,0} - D_2\mu^2 \right)^2 - \Gamma^2 \left( 4 g_\mathrm{opt} g_\mathrm{th} + 3g_\mathrm{opt}^2 - g_\mathrm{th}^2 \right) } }{6g_\mathrm{opt}^2 + 8 g_\mathrm{opt} g_\mathrm{th} + 2g_\mathrm{th}^2}.
\end{aligned}
\end{equation}
This result yields two solutions, which together form an area. To analyze this in detail, the result of equation \ref{equ:critical_power} is presented in Fig. \ref{fig:dispersion} as a function of the pump mode's bare frequency detuning $\Delta_{p,0}$ and for various mode numbers $\mu$. The solution assuming a normal dispersion ring resonator is depicted in the left plot and that for an anomalous dispersion in the right plot. The threshold amplitude $|\alpha_\mathrm{th}|^2$ is converted to optical power, as detailed in Appendix \ref{sec:conversion}. Each differently colored area represents a solution for a distinct mode number $\mu$, starting with $\mu=0$ in dark blue. Additionally, the minimum threshold power $P_\mathrm{th}$ is indicated by the black dashed line and the optical power $P_p$ of the pump mode $\alpha_p$ is shown in red, obtained by solving the steady state of equation \ref{equ:eqm_pump}. The SPM and XPM detunings cause the tilting of the modes and the observed bistability in the pump mode.\\
It is crucial to understand that a classical frequency-comb mode is generated when the red line representing the pump power $P_p$ intersects the line of a possible signal or idler mode $|\alpha_\mathrm{th}|^2$ for a given mode number $\mu$. Obviously, this overlap is easier achievable for a ring resonator with anomalous dispersion. In a normal dispersion ring resonator, the dispersion $D_2$ causes each mode number $\mu$ to have an increasing frequency distance from the pump mode, preventing the lines from ever overlapping. Instead, the frequency detuning for each mode number $\Delta_{\mu}$ increases with the total optical power in the resonator due to SPM and XPM effects. Only with anomalous dispersion, $D_2$ causes the detuning of a mode $\Delta_{\mu}$ to shift towards the pump mode. This phenomenon is well-known in the literature \cite{Herr2012}, which is why classical frequency combs are often generated in ring resonators utilizing anomalous dispersion \cite{FujiiTanabe+2020+1087+1104}.\\
To derive the mode number that first appears in a classical frequency comb, the detuning for each mode is analyzed in detail. Using equation \ref{equ:eqm_pump} and assuming steady-state operation of the pump mode at injection locking with $|\alpha_p|^2=\Gamma P_\mathrm{in}/2g_\mathrm{opt}P_\mathrm{th}$, the detuning of the pump mode can be simplified to
\begin{equation}\label{equ:detuning_p}
\Delta_{p,0} =  -\frac{g_\mathrm{tot}\Gamma}{2g_\mathrm{opt}}\frac{P_\mathrm{in}}{P_\mathrm{th}},
\end{equation}
with the total nonlinearity $g_\mathrm{tot} = g_\mathrm{opt} + g_\mathrm{th}$. The detuning for each mode number $\mu$ can be derived using the same conditions and equation \ref{equ:signalDetuning}, which leads to
\begin{equation}\label{equ:generalDetuning}
\begin{aligned}
	\Delta_\mu = \frac{1}{2} D_2 \mu - \frac{\Gamma\left(g_\mathrm{th} + 2g_\mathrm{opt}\right)}{2g_\mathrm{opt}} \frac{P_\mathrm{in}}{P_\mathrm{th}}.
\end{aligned}
\end{equation}
A frequency comb is generated if the detunings match each other with $\Delta_\mu = \Delta_{p,0}$, which leads to the effective mode detuning for a signal and idler mode as the difference $\Delta_{\mu,\mathrm{eff}} = \Delta_{\mu} - \Delta_{p,0}$ with
\begin{equation}\label{equ:effective_detuning}
\begin{aligned}
\Delta_{\mu, \mathrm{eff}} = \frac{1}{2} D_2 \mu^2 - \frac{\Gamma}{2}\frac{P_\mathrm{in}}{P_\mathrm{th}}.
\end{aligned}
\end{equation}
The mode number of the first appearing frequency-comb mode can be determined by setting $\Delta_{\mu, \mathrm{eff}} = 0$. This leads to the well-known equation
\begin{equation}\label{equ:mode_number}
\begin{aligned}
\mu = \pm \sqrt{\frac{\Gamma}{D_2}\frac{P_\mathrm{in}}{P_\mathrm{th}}}\overset{P_\mathrm{in}=P_\mathrm{th}}{=}\pm \sqrt{\frac{\Gamma}{D_2}},
\end{aligned}
\end{equation}
which has already been derived in \cite{Herr2012}. However, for the quantum description of the frequency comb, the effective detuning $\Delta_{\mu, \mathrm{eff}}$ is crucial to describe the quantum behavior of the individual modes in section \ref{cha:quantumDynamics}. Interestingly, the equations \ref{equ:effective_detuning} and \ref{equ:mode_number} depend only on the optical nonlinearity $g_\mathrm{opt}$ via $P_\mathrm{th}$, while the thermal nonlinearity $g_\mathrm{th}$ cancels out. This is expected, since the influence of the thermal detuning is the same across all modes.

\subsection{Quantum dynamics}
\label{cha:quantumDynamics}
\begin{figure}
	\centering\includegraphics[width=8.6cm]{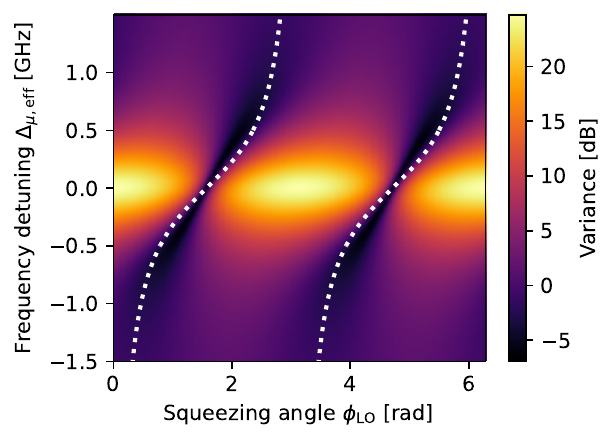}
	\caption{Squeezing spectrum of the generated squeezed light in dependence of the effective frequency detuning $\Delta_{\mu, \mathrm{eff}}$ and the squeezing angle $\phi_{\mathrm{LO}}$ for $\kappa=800$ MHz, $\gamma=200$ MHz, $g_\mathrm{opt}=1.5$ MHz, $\eta=1$ and $\lambda=1550$ nm. The optimum squeezing angle $\phi_{\mathrm{LO},\mathrm{opt}}$ is indicated by the white dotted line.}
	\label{fig:squeezingSpectrum}
\end{figure}
\begin{figure*}
	\centering\includegraphics[width=17.75cm]{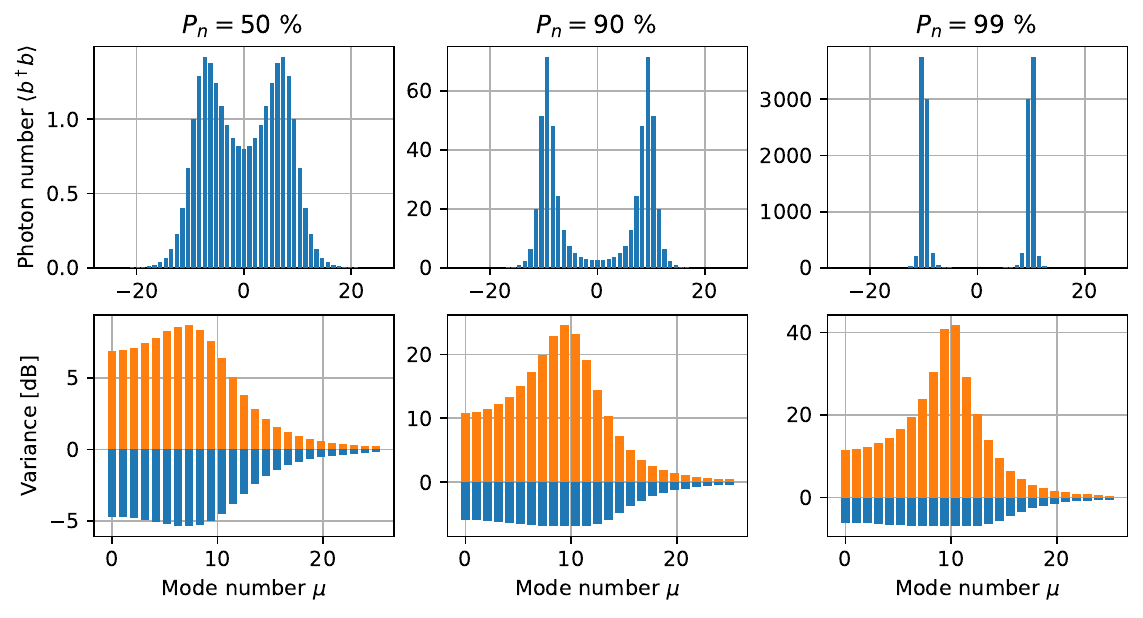}
	\caption{\label{fig:frequencyComb_p} Photon and squeezing spectrum of the quantum frequency comb for an anomalous dispersion ringresonator at the optimum squeezing angle $\phi_{\mathrm{LO},\mathrm{opt}}$. The results are shown for various mode numbers $\mu$, $\kappa=800$ MHz, $\gamma=200$ MHz, $g_\mathrm{opt}=1.5$ MHz, $D_2=10$ MHz, $\eta=1$, $\lambda=1550$ nm and different normalized pump powers $P_n$. The orange values in the squeezing spectrum correspond to the anti-squeezing, while the blue values correspond to the squeezing.}
\end{figure*}

As discussed in the preceding section, a classical frequency comb appears when the pump power $P_\mathrm{in}$ is at least $P_\mathrm{th}$. However, quantum fluctuations and optical modes exhibiting quantum effects are generated even at lower input power $P_\mathrm{in}<P_\mathrm{th}$. These quantum modes are analyzed in detail in the following sections, with a particular focus on the squeezing, the second-order correlation function and the JSI of the quantum frequency comb.\\
Since we are primarily interested in the modes coupled out from the resonator into the straight waveguide to form the waveguide mode $b_\mathrm{out}$, we begin by applying the input-output theory. Therefore, the Hamiltonian of equation \ref{equ:fourWaveMixing_H} is linearized by replacing $a_p$ with the classical mode amplitude $\alpha_p$ and the equations of motion for the out coupled signal and idler modes are derived. This derivation has already been performed in our previous work \cite{tritschler2024optical}, leading to the expression for the out coupled modes of the ring resonator in the frequency domain
\begin{eqnarray}\label{equ:outputMode}
\begin{aligned}
\mathbf{B_{out}}(\omega) = &- \frac{1}{\sqrt{\kappa}}  \Bigl\lbrack[\mathbf{\Omega}-\mathbf{K}-\frac{\kappa}{2} \mathbf{I_4}][\mathbf{\Omega}-\mathbf{K}+\frac{\kappa}{2} \mathbf{I_4}]^{-1} &\\
&\cdot\left( \sqrt{\kappa}\mathbf{B_{in}}(\omega) + \sqrt{\gamma} \mathbf{B_{\gamma}}(\omega) 
\right)- \sqrt{\gamma} \mathbf{B_{\gamma}}(\omega) \Bigr\rbrack, &
\end{aligned}
\end{eqnarray}
where $\mathbf{B_{in}}$ represents the input vector, $\mathbf{B_{\gamma}}$ the loss vector and $\mathbf{B_{out}}$ the output vector and $\mathbf{K}$ includes the geometry, detuning and pump parameters with
\begin{equation}\label{equ:matrix_def}
\begin{aligned}
\mathbf{B_{in}} = \left(\begin{array}{c}
b_{\mathrm{in},s}\\
b_{\mathrm{in},s}^{\dagger} \\
b_{\mathrm{in},i} \\
b_{\mathrm{in},i}^{\dagger}
\end{array}\right), 
\mathbf{B_{\gamma}} = \left(\begin{array}{c}
b_{\gamma, s,}\\
b_{\gamma,s}^{\dagger} \\
b_{\gamma,i} \\
b_{\gamma,i}^{\dagger}
\end{array}\right), 
\mathbf{B_{out}} = \left(\begin{array}{c}
b_{\mathrm{out},s}\\
b_{\mathrm{out},s}^{\dagger} \\
b_{\mathrm{out},i} \\
b_{\mathrm{out},i}^{\dagger}
\end{array}\right),	
\end{aligned}
\end{equation}
\begin{equation}\label{equ:k_matrix}
\begin{aligned}
{\scriptsize
\mathbf{K} = \left(\begin{matrix} 
-i\Delta_{s,\mu, \mathrm{eff}} - \frac{\gamma}{2} & 0 & 0 & \frac{\sigma}{2} \\
0 & i \Delta_{s,\mu, \mathrm{eff}} - \frac{\gamma}{2} & \frac{\sigma^*}{2} & 0 \\
0 & \frac{\sigma}{2} &  - i \Delta_{i,\mu, \mathrm{eff}} - \frac{\gamma}{2} & 0 \\
\frac{\sigma^*}{2} & 0 & 0 & i \Delta_{i,\mu, \mathrm{eff}} - \frac{\gamma}{2} \\
\end{matrix} \right),}
\end{aligned}
\end{equation}
and the pump parameter $\sigma$, which is defined as 
\begin{equation}
	\sigma = 2 g_\mathrm{opt}\alpha_p^2 = \Gamma \frac{P_\mathrm{in}}{P_\mathrm{th}}.
\end{equation}
A key distinction from \cite{tritschler2024optical} is our introduction of the effective mode detunings for the signal ($\Delta_{s,\mu, \mathrm{eff}}$) and idler ($\Delta_{i,\mu, \mathrm{eff}}$) modes in equation \ref{equ:k_matrix}, which enables the calculation of the different mode numbers. \\
Equation \ref{equ:outputMode} describes the output modes, including losses within the ring resonator. However, after coupling out to the straight waveguide, these modes are subject to additional losses before being utilized for specific operations. We model these post coupling losses using the standard procedure involving a beam splitter, which combines the output modes with vacuum modes according to $\sqrt{\eta}\mathbf{B_{out}}(\omega) + \sqrt{1-\eta}\mathbf{B_v}$. Here the vacuum vector is defined as $\mathbf{B_v}=[b_v, b_v^\dagger]^\top$ and $\eta$ denotes the efficiency.\\
Since the modes in $\mathbf{B_v}$, $\mathbf{B_{in}}$ and $\mathbf{B_{\gamma}}$ correspond to vacuum modes when only one mode at $\omega_p$ is used to pump the ring resonator mode $a_p$, the expectation values of the output modes can be obtained by utilizing equation \ref{equ:outputMode} and the well-known expectation values for vacuum modes $\langle b_v( t) \rangle = \langle b_v^{\dagger}(t) \rangle = \langle b_v( t) b_v( t' ) \rangle = \langle b_v^{\dagger}( t) b_v( t' ) \rangle = 0$ and $\langle b_v\left( t \right) b_v^{\dagger}( t' ) \rangle = \delta(t-t')$ \cite{zoller1997quantum, wiseman_milburn_2009}. It is important to note that due to linearization, equation \ref{equ:outputMode} is valid only below the FWM threshold, specifically for an input power $P_\mathrm{in}$ up to $99.895 \%$ of $P_\mathrm{th}$ \cite{tritschler2024optical}.\\
Assuming a symmetric behavior for the signal and idler modes with $\Delta_{s,\mu, \mathrm{eff}} = \Delta_{i,\mu, \mathrm{eff}} = \Delta_{\mu, \mathrm{eff}}$, equation \ref{equ:outputMode} can be used to derive, for example, the photon number of the fluctuating signal mode of the mode number $\mu$ to
\begin{equation}\label{equ:photon_n}
\begin{aligned}
\langle b_{\mathrm{out},s}^\dagger \left(\omega \right) b_{\mathrm{out},s} \left(\omega' \right) \rangle = \frac{4\Gamma^3\kappa \eta P_\mathrm{in}^2 P_\mathrm{th}^2 }{\left( 4\Delta_{\mu, \mathrm{eff}}^2 P_\mathrm{th}^2 + \Gamma^2\left[ P_\mathrm{th}^2 - P_\mathrm{in}^2 \right] \right)^2} \delta'.
\end{aligned}
\end{equation}
In the same way, the expectation values of each mode $\mu$ in the quantum frequency comb can be determined. \\
Finally, we can use the results obtained so far to analyze the quantum behavior of the generated fluctuations. We begin by examining the squeezing between the signal and idler modes generated via FWM. For this purpose, the two-mode quadrature operator is introduced as
\begin{equation} \label{equ:quadrature_signal}
X_Q = \frac{1}{\sqrt{2}}\left[ (b_{\mathrm{out},s} + b_{\mathrm{out},i} ) e^{i \phi_\mathrm{LO}} + (b_{\mathrm{out},s}^{\dagger}  + b_{\mathrm{out},i}^{\dagger}) e^{-i \phi_\mathrm{LO}}\right]
\end{equation}
and with the local oscillator phase $\phi_{\mathrm{opt}}$, which corresponds to the squeezing angle \cite{Walls2008}. To analyze the squeezing, the variance of $X_Q$ is calculated with
\begin{equation}\label{equ:variance}
\begin{aligned}
\langle V\left(\omega,\omega'\right) \rangle &= \Delta X_Q\left(\omega,\omega'\right) \\
&= \langle X_Q\left(\omega\right) X_Q\left(\omega'\right) \rangle - \langle X_Q\left(\omega\right) \rangle\langle X_Q\left(\omega'\right) \rangle \\
&\xrightarrow{b_{\mathrm{in},s/i}=b_v} \langle X_Q\left(\omega\right) X_Q\left(\omega'\right) \rangle.
\end{aligned}
\end{equation}
For simplicity, we assume a symmetric behavior between the generated signal and idler mode with $\langle b_\mathrm{out,s}^\dagger b_\mathrm{out,s}\rangle \approx \langle b_\mathrm{out,i}^\dagger b_\mathrm{out,i}\rangle$ and $\langle b_\mathrm{out,i} b_\mathrm{out,s}\rangle \approx \langle b_\mathrm{out,s} b_\mathrm{out,i}\rangle$. This leads to the simplified result of the variance with
\begin{eqnarray}\label{equ:variance_general1}
\begin{aligned}
\langle V(\omega, \omega')\rangle = &\langle b_\mathrm{out,s}b_\mathrm{out,i}\rangle e^{2i\phi_{\mathrm{LO}}}+\langle b_\mathrm{out,s}^\dagger b_\mathrm{out,i}^\dagger\rangle e^{-2i\phi_{\mathrm{LO}}} \\ &+ 2\langle b_\mathrm{out,s}^\dagger b_\mathrm{out,s}\rangle + 1
\end{aligned}
\end{eqnarray}
using the expectation value of the photon number from equation \ref{equ:photon_n} and 
\begin{equation}\label{equ:bsbi_exp}
\begin{aligned}
\langle b_\mathrm{out,s}b_\mathrm{out,i} \rangle = \frac{2\kappa\Gamma \eta P_\mathrm{in} P_\mathrm{th}\left( \Gamma^2 P_\mathrm{in}^2 - P_\mathrm{th}^2 \left[ 2\Delta_{\mu, \mathrm{eff}} + i \Gamma \right]^2 \right)}{\left( 4\Delta_{\mu, \mathrm{eff}}^2 P_\mathrm{th}^2 + \Gamma^2\left[ P_\mathrm{th}^2 - P_\mathrm{in}^2 \right] \right)^2}.
\end{aligned}
\end{equation}
Note that the expectation value of $\langle b_\mathrm{out,s}^\dagger b_\mathrm{out,i}^\dagger\rangle$ corresponds to the complex conjugate of $\langle b_\mathrm{out,s}b_\mathrm{out,i} \rangle$. \\
To analyze this result, the squeezing spectrum is presented in Fig. \ref{fig:squeezingSpectrum}. A dominant squeezing of approximately $-5.3$ dB and an anti-squeezing of approximately $21$ dB are observed at $\Delta_{\mu, \mathrm{eff}}=0$. It is crucial to understand that the squeezing behavior for any other mode number $\mu$ can be determined simply by using its corresponding effective detuning $\Delta_{\mu, \mathrm{eff}}$. Additionally, it is interesting that the optimal squeezing angle $\phi_{\mathrm{LO}}$ depends on the effective detuning $\Delta_{\mu, \mathrm{eff}}$ and, consequently, on $\mu$. Ideally, the optimum squeezing occurs at an angle of $\phi_{\mathrm{LO}}=n(\pi/2)$ with $n=1,3,\ldots.$ and anti-squeezing at $\phi_{\mathrm{LO}}=n\pi$ with $n=0,1,2,\ldots.$, as it is well known in the literature. However, when generating a complete quantum frequency comb, this ideal scenario can only be achieved for a few signal and idler mode pairs. This limitation arises because it is impossible to attain $\Delta_{\mu, \mathrm{eff}}=0$ for every mode $\mu$ simultaneously, due to the combined effects of XPM, SPM, and the ring dispersion. \\
\begin{figure*}
	\centering\includegraphics[width=17.75cm]{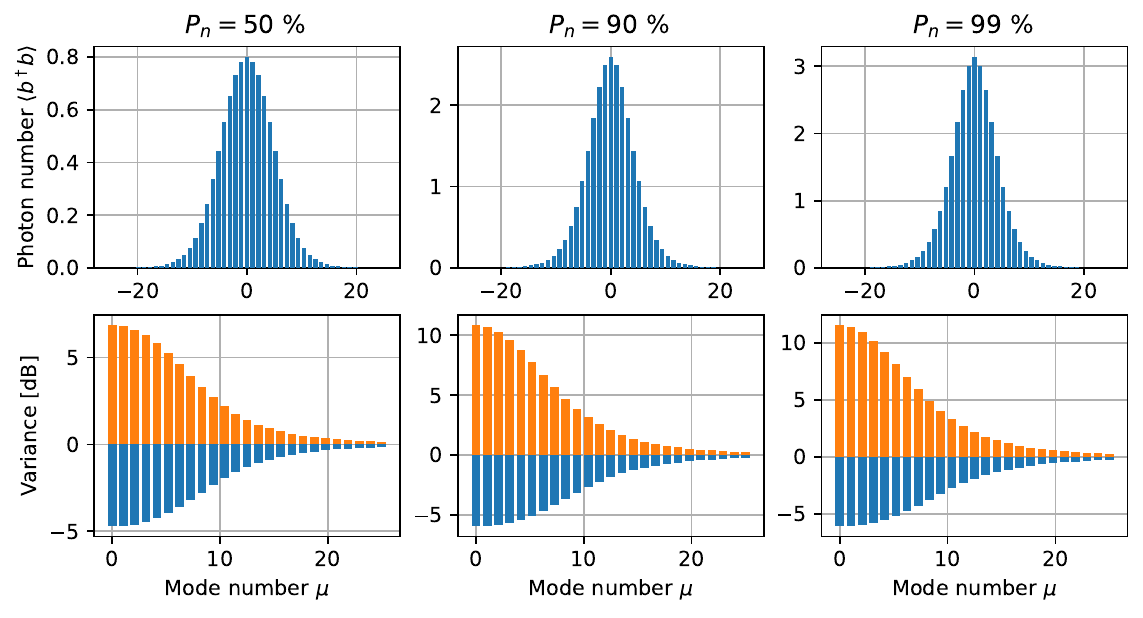}
	\caption{\label{fig:frequencyComb_n} Photon and squeezing spectrum of the quantum frequency comb for a normal dispersion ringresonator at the optimum squeezing angle $\phi_{\mathrm{LO},\mathrm{opt}}$. The results are shown for various mode numbers $\mu$, $\kappa=800$ MHz, $\gamma=200$ MHz, $g_\mathrm{opt}=1.5$ MHz, $D_2=-10$ MHz, $\eta=1$, $\lambda=1550$ nm and different normalized pump powers $P_n$. The orange values in the squeezing spectrum correspond to the anti-squeezing, while the blue values correspond to the squeezing.}
\end{figure*}
Using equation \ref{equ:variance_general1}, we can determine the optimal squeezing angle for each mode number $\mu$ to achieve the best squeezing as
\begin{equation}\label{equ:optimal_phase}
\phi_{\mathrm{LO},\mathrm{opt}}=\begin{cases}
-\frac{1}{2}\tan^{-1}\left(\frac{4\Delta_{\mu, \mathrm{eff}} \Gamma}{4\Delta_{\mu, \mathrm{eff}}^2 -\Gamma^2 \left[1 - \left( \frac{P_\mathrm{in}}{P_\mathrm{th}} \right)^2\right]}\right), \\ \qquad\qquad\qquad\qquad\qquad \text{$4\Delta_\mu^2 -\Gamma^2 - \sigma^2>0$},\\
-\frac{1}{2}\tan^{-1}\left(\frac{4\Delta_{\mu, \mathrm{eff}} \Gamma}{4\Delta_{\mu, \mathrm{eff}}^2 -\Gamma^2 \left[1 - \left( \frac{P_\mathrm{in}}{P_\mathrm{th}} \right)^2\right]}\right) + \frac{\pi}{2},\\ \qquad\qquad\qquad\qquad\qquad \text{$4\Delta_\mu^2 -\Gamma^2 - \sigma^2\leq0$}.
\end{cases}
\end{equation}
The result of $\phi_{\mathrm{LO},\mathrm{opt}}$ is sketched as the white dashed line in Fig. \ref{fig:squeezingSpectrum}. Obviously, the squeezing angle of the anti-squeezing is detuned by $\pi/2$. This result is used to simplify the equations for the squeezing and anti-squeezing at $\phi_{\mathrm{LO},\mathrm{opt}}$ to 
\begin{equation}\label{equ:tm_squeezing}
\small
\begin{aligned}
\frac{\langle V_\mathrm{s} \rangle }{\langle V_\mathrm{vac} \rangle} &= 1 + 2\langle b_\mathrm{out,s}^\dagger b_\mathrm{out,s} \rangle - 2\left| \sqrt{\langle b_\mathrm{out,s} b_\mathrm{out,i} \rangle}  \right|^2 \\
&= 1 + \frac{8\eta\kappa\Gamma^3}{\left(4\Delta_{\mu, \mathrm{eff}}^2 + \Gamma^2 \left[1 - \frac{P_\mathrm{in}}{P_\mathrm{th}}\right]\right)} \\
&- 4 \left| \sqrt{ \frac{\eta\kappa\Gamma P_\mathrm{in} P_\mathrm{th}\left( \Gamma^2 P_\mathrm{in}^2 - P_\mathrm{th}^2 \left[ 2\Delta_{\mu, \mathrm{eff}} + i \Gamma \right]^2 \right)}{\left( 4\Delta_{\mu, \mathrm{eff}}^2 P_\mathrm{th}^2 + \Gamma^2\left[ P_\mathrm{th}^2 - P_\mathrm{in}^2 \right] \right)^2}} \right|^2,
\end{aligned}
\end{equation}
\begin{equation}\label{equ:tm_AntiSqueezing}
\small
\begin{aligned}
\frac{\langle V_\mathrm{as} \rangle }{\langle V_\mathrm{vac} \rangle} &= 1 + 2\langle b_\mathrm{out,s}^\dagger b_\mathrm{out,s} \rangle + 2\left| \sqrt{\langle b_\mathrm{out,s} b_\mathrm{out,i} \rangle}  \right|^2 \\
&= 1 + \frac{8\eta\kappa\Gamma^3}{\left(4\Delta_{\mu, \mathrm{eff}}^2 + \Gamma^2 \left[1 - \frac{P_\mathrm{in}}{P_\mathrm{th}}\right]\right)} \\
&+ 4 \left| \sqrt{ \frac{\eta\kappa\Gamma  P_\mathrm{in} P_\mathrm{th}\left( \Gamma^2 P_\mathrm{in}^2 - P_\mathrm{th}^2 \left[ 2\Delta_{\mu, \mathrm{eff}} + i \Gamma \right]^2 \right)}{\left( 4\Delta_{\mu, \mathrm{eff}}^2 P_\mathrm{th}^2 + \Gamma^2\left[ P_\mathrm{th}^2 - P_\mathrm{in}^2 \right] \right)^2}} \right|^2.
\end{aligned}
\end{equation}
The derived equations simplify to the results presented in \cite{tritschler2025chipintegratedsqueezedlight} for the case of $\mu=0$, validating our equations. Neglecting the detuning with $\Delta_{\mu, \mathrm{eff}}=0$, the equations for the optimum squeezing and anti-squeezing can also be derived as
\begin{equation}\label{equ:tm_squeezing_opt}
\begin{aligned}
\frac{\langle V_\mathrm{s,\mathrm{opt}} \rangle }{\langle V_\mathrm{vac} \rangle} = 1 - \frac{4\kappa\eta P_\mathrm{in}P_\mathrm{th}}{\Gamma\left( P_\mathrm{in} + P_\mathrm{th} \right)} \xrightarrow[]{P_\mathrm{in}\rightarrow P_\mathrm{th}} 1-\frac{\eta\kappa}{\Gamma},
\end{aligned}
\end{equation}
\begin{equation}\label{equ:tm_AntiSqueezing_opt}
\begin{aligned}
\frac{\langle V_\mathrm{as,\mathrm{opt}} \rangle }{\langle V_\mathrm{vac} \rangle} = 1 + \frac{4\kappa\eta P_\mathrm{in}P_\mathrm{th}}{\Gamma\left( P_\mathrm{in} - P_\mathrm{th} \right)} \xrightarrow[]{P_\mathrm{in}\rightarrow P_\mathrm{th}} \infty.
\end{aligned}
\end{equation}
The derived results for the photon number of the fluctuations, as well as the squeezing and anti-squeezing, are presented in Fig. \ref{fig:frequencyComb_p} for anomalous dispersion at various input powers, indicated by the normalized pump power $P_n=P_\mathrm{in}/P_\mathrm{th}$. It is important to note that the mode at $\mu=0$ corresponds to a single-mode squeezed state, as the generated photons share the same angular frequency as the pump, $\omega_p$. All other modes with $\mu\neq0$ represent a generated signal and idler pair and thus a two-mode squeezed state. At a low input power of $P_n=50$ \%, many modes are stimulated with a rather low photon number. As $P_n$ increases, the photon number also rises, becoming significant only for a few modes that exhibit the smallest effective detuning $\Delta_{\mu, \mathrm{eff}}$. While both squeezing and anti-squeezing demonstrably increase with $P_n$, the squeezing is already significant for multiple modes even at a relatively low input power $P_n$. Therefore, a wide range of modes exhibit significant squeezing, even though only a few of them consist of a large photon number.\\
\begin{figure}
	\centering\includegraphics[width=8.6cm]{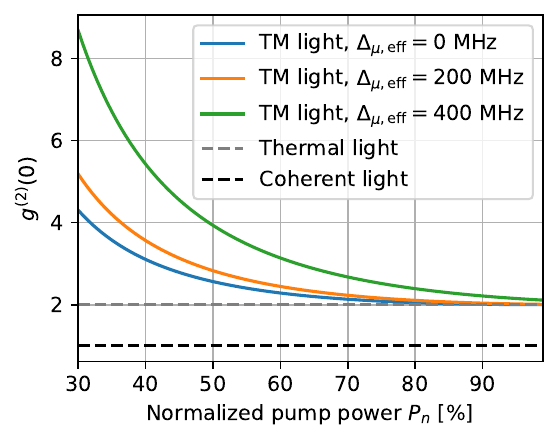}
	\caption{Second order correlation function $g^{(2)}(0)$ of coherent light (black dashed line), thermal light (orange dashed line) and two-mode light at different effective detuning values with $\Delta_{\mu, \mathrm{eff}}=0$ MHz (blue line), $\Delta_{\mu, \mathrm{eff}}=200$ MHz (orange line) and $\Delta_{\mu, \mathrm{eff}}=400$ MHz (green line) in dependency of the normalized pump power $P_n$. The signal and idler modes respectively correspond to thermal light, while both combined form the two-mode light. All values are shown for $\kappa=800$ MHz, $\gamma=200$ MHz, $g_\mathrm{opt}=1.5$ MHz and $\lambda=1550$ nm.}
	\label{fig:g2_function}
\end{figure}
The case of a normal dispersion ring resonator is also very interesting and presented in Figure \ref{fig:frequencyComb_n}. From the classical results in Fig. \ref{fig:dispersion}, it is known that the FWM threshold power can never be achieved for any mode number $\mu$ due to normal dispersion, thus preventing the generation of a classical frequency comb. This aligns with the photon number results shown in Fig. \ref{fig:frequencyComb_n}, which remain very small across all $P_n$ values. The $\mu=0$ mode exhibits the largest photon number, as it corresponds to the smallest $\Delta_{\mu, \mathrm{eff}}$. Interestingly, significant squeezing is still observed for a few modes, though the overall squeezing remains smaller compared to that achieved with an anomalous dispersion ring resonator. Nevertheless, this is an important finding, as it demonstrates that a normal dispersion ring resonator can also be utilized to generate a quantum frequency comb. This could be particularly valuable if a single-mode squeezed state is desired, given that the squeezing results for the $\mu=0$ mode are identical in the normal and anomalous regime.\\
\begin{figure*}
	\includegraphics[width=8.5cm]{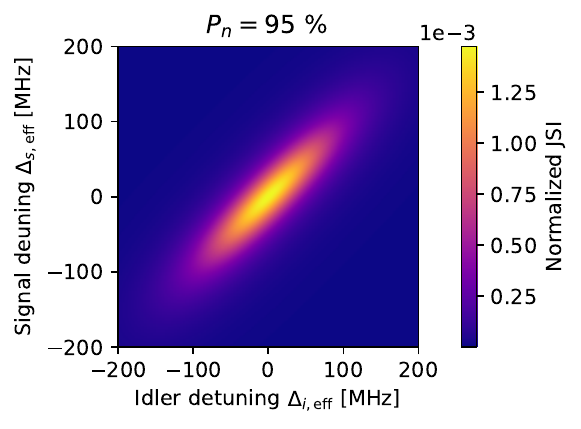}
	\includegraphics[width=8.5cm]{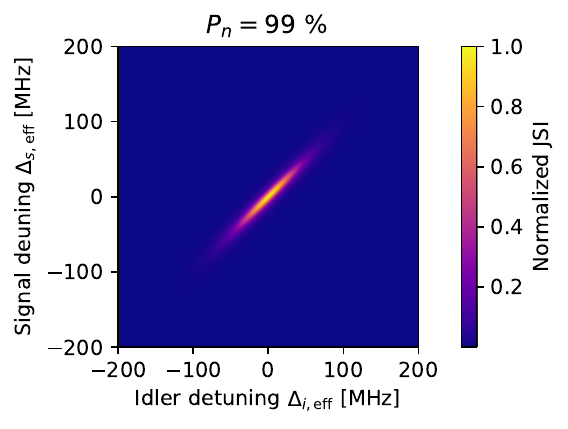}
	\caption{\label{fig:jsi} JSI in dependence of the effective detuning of the signal and idler mode $\Delta_{s,\mu, \mathrm{eff}}$ and  $\Delta_{i,\mu, \mathrm{eff}}$ at different normalized pump power values $P_n$. All values are shown for $\kappa=800$ MHz, $\gamma=200$ MHz, $g_\mathrm{opt}=1.5$ MHz and $\lambda=1550$ nm.}
\end{figure*}
Following the analysis of squeezing in the quantum frequency comb, we investigate the correlation between the modes. The second-order correlation function serves as an indicator for the particle correlation, quantifying the two-photon correlation between a generated signal and idler photon. It can be determined using
\begin{equation}
	g^{(2)}(t) = \frac{\langle b_\mathrm{out,s}^\dagger(t) b_\mathrm{out,i}^\dagger(t) b_\mathrm{out,s}(t) b_\mathrm{out,i}(t)\rangle}{\langle b_\mathrm{out,s}^\dagger(t) b_\mathrm{out,s}(t)\rangle\langle b_\mathrm{out,i}^\dagger(t) b_\mathrm{out,i}(t)\rangle}.
\end{equation}
We evaluate $g^{(2)}_s(t)$ at $t=0$, which is of interest in most applications. Note that the higher-order expectation values are first reduced to second order using the cumulant expansion from \cite{Kubo1962}, and then solved using the expectation values derived with equation \ref{equ:outputMode}. The result for the signal and idler modes, respectively, is $g^{(2)}_s(0)=2$, indicating that both the signal and idler modes correspond to a thermal state. In comparison, the joint second-order correlation between the signal and idler mode is given by
\begin{equation}
\small
g^{(2)}_{si}(0) = \frac{1}{4}\left( \left[ \frac{\left(4\Delta_{\mu, \mathrm{eff}}^2 + \Gamma^2 \right)}{\Gamma^2} \frac{P_\mathrm{th}}{P_\mathrm{in}} \right]^2 + \left[\frac{P_\mathrm{in}}{P_\mathrm{th}} \right]^2 - \frac{8\Delta_{\mu, \mathrm{eff}}^2}{\Gamma^2} + 6 \right).
\end{equation}
The results are presented in Figure \ref{fig:g2_function} for various values of $\Delta_{\mu, \mathrm{eff}}$ as a function of the normalized pump power $P_n$. Additionally, the values for thermal and coherent light, $g^{(2)}_{\mathrm{thermal}}(0) = 2$ and $g^{(2)}_{\mathrm{coherent}}(0)=1$ respectively, are also shown. It can be observed that $g^{(2)}_{si}(0)$ increases with $\Delta_{\mu, \mathrm{eff}}$ and decreases with increasing $P_n$ with converging to a thermal state as $P_\mathrm{in} \rightarrow P_\mathrm{th}$. The reason for this behavior is that at low pump power $P_n$, only a few signal and idler photons are generated. However, if a signal photon is generated, there is a high probability that the corresponding idler photon is the particle-entangled partner. For higher input power, a larger number of photons are generated, leading to a reduction in the correlation between individual signal and idler photons. \\
Besides two-photon correlation, spectral mode correlation is another important feature analyzed in this work. It can be assessed using the JSI, which represents the joint signature probability distribution and indicates the probability of detecting a signal mode at a certain frequency if an idler mode is detected at another frequency. Thus, it is a value describing the correlation between the two modes as a function of frequency or time \cite{Zielnicki2018}.\\
For the derivation of the JSI, the effective detunings of the signal and idler modes are considered individually and thus the JSI can be determined as
\begin{widetext}
\begin{equation}\label{equ:jsi}
\begin{aligned}
\Phi &= \langle b_\mathrm{out,s}^\dagger(\omega) b_\mathrm{out,i}^\dagger(\omega) b_\mathrm{out,s}(\omega) b_\mathrm{out,i}(\omega)\rangle \\ &= \frac{4 \kappa^2\Gamma^2 P_\mathrm{in}^2 P_\mathrm{th}^2 \left( \Gamma^4 P_\mathrm{in}^4 + 2 \Gamma^2 P_\mathrm{in}^2 P_\mathrm{th}^2 \left[ 4\Delta_i \Delta_s + 3 \Gamma^2 \right]  + P_\mathrm{th}^4 \left[ 4\Delta_i^2 + \Gamma^2 \right] \left[ 4\Delta_s^2 + \Gamma^2 \right] \right) }{\left( \Gamma^4 P_\mathrm{in}^4 + 2 \Gamma^2 P_\mathrm{in}^2 P_\mathrm{th}^2 \left[ 4\Delta_i \Delta_s - \Gamma^2 \right]  + P_\mathrm{th}^4 \left[ 4\Delta_i^2 + \Gamma^2 \right] \left[ 4\Delta_s^2 + \Gamma^2 \right] \right)^2}.
\end{aligned}
\end{equation}
\end{widetext}
The result of equation \ref{equ:jsi} is shown in Fig. \ref{fig:jsi} for normalized pump powers of $P_n=95$ \% and $P_n=99$ \%. The plotted values are normalized by the maximum JSI value of 125396342.22. It can be observed that the JSI increases with $P_n$ and decreases significantly with an asymmetric detuning of the signal or idler mode, while exhibiting greater stability for symmetric detuning. This behavior can be explained with the energy and momentum conservation of the FWM process $\omega_s + \omega_i = 2\omega_p$. Consequently, if the signal mode is detuned by a certain amount, the idler mode must undergo an inverse frequency shift to satisfy the conservation relation $\omega_s + \omega_i = 2\omega_p$.

\subsection{Entanglement}

\begin{figure}
	\centering\includegraphics[width=8.6cm]{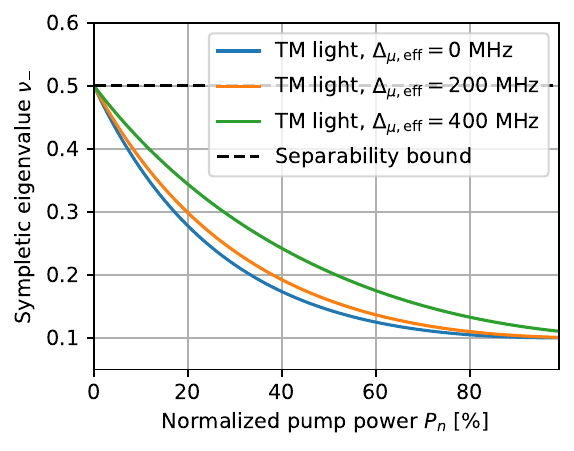}
	\caption{Smallest sympletic eigenvalue $v_-$ of two-mode light at different effective detuning values with $\Delta_{\mu, \mathrm{eff}}=0$ MHz (blue line), $\Delta_{\mu, \mathrm{eff}}=200$ MHz (orange line) and $\Delta_{\mu, \mathrm{eff}}=400$ MHz (green line) in dependence of the normalized pump power $P_n$. The separability bound is indicated by the black dashed line and all values $v_-<1/2$ violate it, showing entanglement. All values are shown for $\kappa=800$ MHz, $\gamma=200$ MHz, $g_\mathrm{opt}=1.5$ MHz and $\lambda=1550$ nm.}
	\label{fig:entanglement}
\end{figure}

\begin{figure}
	\centering\includegraphics[width=8.6cm]{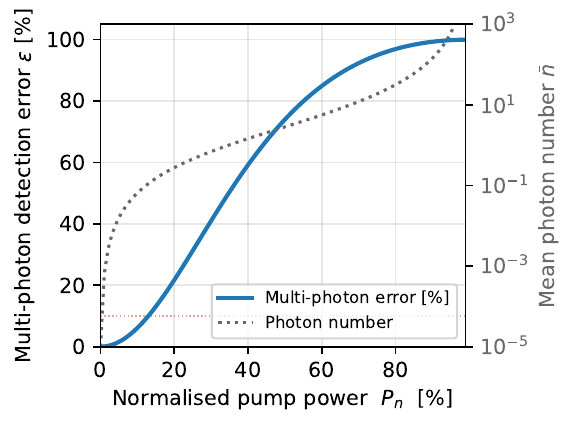}
	\caption{Multi-photon detection error rate (blue solid line), which rises with the photon number of the generated signal and idler mode (gray dotted line). }
	\label{fig:entanglement_error}
\end{figure}

Finally, the last investigated quantum feature of frequency combs is the most famous one with the entanglement, which is important for many quantum information and computating protocols. While $g^{\left(2\right)}$ and the JSI analyze the correlation between the photon pairs and modes, the entanglement between the signal and idler mode is investigated using the standard Simon criterion for continous variable quantum systems \cite{Simon_2000}. \\
Therefore, the canonical operators of the signal and idler modes are defined first
\begin{equation}\label{equ:s_i_quadrature}
	\begin{aligned}
		Q_{s,i} = \frac{b_\mathrm{out,s}+b_\mathrm{out,s}^\dagger}{\sqrt{2}}, 
		P_{s,i} = \frac{b_\mathrm{out,s}-b_\mathrm{out,s}^\dagger}{i\sqrt{2}},
	\end{aligned}
\end{equation}
which correspond the quadrature operators of the individual modes and are used to calculate the covariance matrix of the modes with
\begin{equation}\label{equ:tm_covariance_matrix}
	\begin{aligned}
 	\mathbf{C} &= {\scriptsize
 	\left(\begin{matrix} 
 		C_{A} & C_{C}  \\
 		C_{C}^T & C_{B}
 	\end{matrix} \right)}& \\
 &= {\scriptsize
 	\left(\begin{matrix} 
 		\langle \left\lbrace Q_s, Q_s  \right\rbrace  \rangle & \langle \left\lbrace Q_s, P_s  \right\rbrace  \rangle & \langle \left\lbrace Q_s, Q_i  \right\rbrace  \rangle & \langle \left\lbrace Q_s, P_i  \right\rbrace  \rangle  \\
 		\langle \left\lbrace P_s, Q_s  \right\rbrace  \rangle & \langle \left\lbrace P_s, P_s  \right\rbrace  \rangle & \langle \left\lbrace P_s, Q_i  \right\rbrace  \rangle & \langle \left\lbrace P_s, P_i  \right\rbrace  \rangle  \\
 		\langle \left\lbrace Q_i, Q_s  \right\rbrace  \rangle & \langle \left\lbrace Q_i, P_s  \right\rbrace  \rangle & \langle \left\lbrace Q_i, Q_i  \right\rbrace  \rangle & \langle \left\lbrace Q_i, P_i  \right\rbrace  \rangle  \\
 		\langle \left\lbrace P_i, Q_s  \right\rbrace  \rangle & \langle \left\lbrace P_i, P_s  \right\rbrace  \rangle & \langle \left\lbrace P_i, Q_i  \right\rbrace  \rangle & \langle \left\lbrace P_i, P_i  \right\rbrace  \rangle  \\
 	\end{matrix} \right)}.&
\end{aligned}
\end{equation}
The operators $Q_{s,i}$ and $P_{s,i}$ can be combined in a vector $\mathbf{X}=(Q_s, P_s, Q_i, P_s)$ with the canonical commutation relation in the sympletic form $[X_i, X_j] = i\Omega_{i,j}$ with
\begin{equation}\label{equ:entanglement_matrix}
	\begin{aligned}
		\mathbf{\Omega} &= {\scriptsize
			\left(\begin{matrix} 
				\omega & 0  \\
				0 & \omega
			\end{matrix} \right), \qquad\omega = \left(\begin{matrix} 
			0 & 1  \\
			-1 & 0
		\end{matrix} \right)}. & 
	\end{aligned}
\end{equation}
As in the previous sections, a symmetric behavior between the signal and the idler mode is assumed and the resulting non vanishing expectation values of $\mathbf{C}$ are
\begin{equation}
\begin{aligned}
	\langle \left\lbrace Q_s, Q_s  \right\rbrace  \rangle &= \langle \left\lbrace P_s, P_s  \right\rbrace  \rangle = \frac{1}{2} + \langle b_\mathrm{out,s}^\dagger b_\mathrm{out,s} \rangle &\\
	\langle \left\lbrace Q_s, Q_i \right\rbrace  \rangle &= \frac{1}{2} \left( \langle b_\mathrm{out,i}^\dagger b_\mathrm{out,s}^\dagger \rangle + \langle b_\mathrm{out,i} b_\mathrm{out,s} \rangle \right) &\\
	\langle \left\lbrace Q_s, P_i  \right\rbrace  \rangle &= \langle \left\lbrace P_s, Q_i  \right\rbrace  \rangle = \frac{i}{2} \left( \langle b_\mathrm{out,i}^\dagger b_\mathrm{out,s}^\dagger \rangle - \langle b_\mathrm{out,i} b_\mathrm{out,s} \rangle \right)& \\
	\langle \left\lbrace P_s, P_i \right\rbrace  \rangle &= - \frac{1}{2} \left( \langle b_\mathrm{out,i}^\dagger b_\mathrm{out,s}^\dagger \rangle + \langle b_\mathrm{out,i} b_\mathrm{out,s} \rangle \right) &\\
	\langle \left\lbrace Q_i, Q_i  \right\rbrace  \rangle &= 	\langle \left\lbrace P_i, P_i  \right\rbrace  \rangle = \frac{1}{2} + \langle b_\mathrm{out,i}^\dagger b_\mathrm{out,i} \rangle &\\
\end{aligned}
\end{equation}
with $\langle \left( a_s^\dagger\right)^2 \rangle = \langle a_s^2  \rangle = 0$ and the already derived results for the expectation values in equations \ref{equ:photon_n} and \ref{equ:bsbi_exp}. Note that the matrix is symetric with $C_C$ due to the symmetry between the signal and the idler mode.\\
The entanglement between two states or modes is present if the two modes are not separable and this indicates a strong correlation between both modes. Simon's positive partial transpose (PPT) criterion states that if a state is separable, the partially transposed counterpart of $\mathbf{C}$, which is determined by $\tilde{\mathbf{C}} = \Lambda \mathbf{C} \Lambda $ and $\Lambda = \mathrm{diag}\left( 1, 1, 1, -1 \right)$, is also separable \cite{Simon_2000, Peres_1996, Horodecki_1997}. It has been shown that this separability can be analyzed by the calculation of the smallest symplectic eigenvalue of $\mathbf{C}$, which is performed using the invariants of Serfaini et al. \cite{Serafini_2003} with
\begin{equation}
\begin{aligned}
	v_{\pm} = \sqrt{\frac{\Delta\left( \mathbf{C} \right) \pm \sqrt{\Delta\left( \mathbf{C} \right)^2 - 4 \mathrm{det}\left(\mathbf{C}\right)}}{2}},
\end{aligned}	
\end{equation}
\begin{equation}
	\begin{aligned}
	\Delta \left( \mathbf{C} \right) &= \mathrm{det}\left(C_{A}\right) + \mathrm{det}\left(C_{B}\right) - 2 \mathrm{det}\left(C_{C}\right)& \\
	&= \frac{1}{2} + 2 \langle b_\mathrm{out,i}^\dagger  b_\mathrm{out,s}^\dagger \rangle \langle b_\mathrm{out,i} b_\mathrm{out,s} \rangle& \\
	&+ 2 \langle b_\mathrm{out,s}^\dagger  b_\mathrm{out,s} \rangle^2 + 2 \langle b_\mathrm{out,s}^\dagger  b_\mathrm{out,s} \rangle& 
	\end{aligned}	
\end{equation}
and
\begin{equation}
	\begin{aligned}
		\mathrm{det}\left( \mathbf{C} \right)  = &    \Bigl( \langle b_\mathrm{out,i}^\dagger  b_\mathrm{out,s}^\dagger \rangle \langle b_\mathrm{out,i}  b_\mathrm{out,s} \rangle &\\
		& - \langle b_\mathrm{out,s}^\dagger  b_\mathrm{out,s} \rangle^2- \langle b_\mathrm{out,s}^\dagger  b_\mathrm{out,s} \rangle - \frac{1}{4}\Bigr)^2.&
	\end{aligned}	
\end{equation}
In the investigated use case, the generated signal and idler modes are entangled if the Heisenberg uncertainty condition $\mathbf{C}+ i\Omega /2 \geq 1/2 $ is violated and thus if the separability condition $v_- \geq 1/2$ is violated \cite{Weedbrook_2012, Serafini_2003}. Note that all our results use the convention $[Q, P]=i$, the vacuum variance equal to $1/2$ and $\hbar =1$. To convert this to the shot-noise units with a vacuum variance of 1, it is required to multiply all elements of $\mathbf{C}$ and the symplectic eigenvalues by 2 and shift the separability threshold to $v_->1$ \cite{Simon_2000, Weedbrook_2012, Serafini_2003}.\\
The results for $v_-$ are shown in Figure \ref{fig:entanglement} and it can be seen that with a rising pump power $v_-$ reduces, the separability is violated at $P_n>0$. Thus, the entanglement increases as expected with $P_n$ and $v_-$ converges to
\begin{equation}
	\begin{aligned}
		v_{-, \mathrm{min}} \xrightarrow[]{P_\mathrm{in}\rightarrow P_\mathrm{th}} \frac{\gamma}{2\Gamma}.
	\end{aligned}	
\end{equation}
This results in the requirement of $\kappa \gg \gamma$ and thus low losses are required to achieve a strong entanglement.\\
For an efficient utilization of the entanglement, it is crucial to choose the best-suited detection method. The common methods measure either the mode entanglement, the entanglement between the signal and idler mode, or the particle entanglement, the entanglement between the signal and the idler photon pair. While for the mode entanglement an intensity measurement at a certain frequency is performed, particle-entanglement measurements require a particle resolution. The intensity measurements can be performed efficient using a homodyne or heterodyne measurement. However, particle entanglement detection measurements are dominated by the multi-photon detection error, which can be given as
\begin{equation}
	\varepsilon = \frac{\bar{n}}{(\bar{n} + 1)},
\end{equation}
with the photon number of the generated signal and idler modes $\bar{n} = \langle a_s^\dagger a_s \rangle + \langle a_i^\dagger a_i \rangle$ \cite{Gisin_2002}. The error rate $\varepsilon$ is shown in Figure \ref{fig:entanglement_error} together with the photon number $\bar{n}$. Generally, only a small $\varepsilon$ is tolerated and thus particle entanglement measurements are only used at very small values for $P_n$, while the best operation to exploit mode entanglement is at $P_\mathrm{in}\approx P_\mathrm{th}$, as it is well known in the literature \cite{Yang2021, Guidry2022, Strekalov_2016, PhysRevLett.114.050501, PhysRevLett.108.083601, PhysRevLett.101.130501, Lu2019, Arrazola2021}

\section{Discussion}
\label{sec:discussion}

In this work we derived closed-form analytical expressions describing the squeezing, second-order correlation, the JSI and the entanglement for each mode within a quantum frequency comb. These quantities are predominantly employed in typical quantum technology applications. Squeezing, for instance, is frequently utilized in sensing applications. Our results demonstrate that significant single- and two-mode squeezing can be generated in both normal and anomalous dispersion regimes. While dispersion is irrelevant for single-mode squeezed light, optimal two-mode squeezing is exclusively achieved under anomalous dispersion, as only then $\Delta_{\mu, \mathrm{eff}}=0$ is possible. If the squeezing of multiple modes in the quantum frequency comb is utilized, it is important to note that each mode corresponds to a different squeezing angle as shown in Fig. \ref{fig:squeezingSpectrum}. This difference can be critical for applications where modes are mixed, as matching the squeezing angles of various modes is essential to enhance or reduce the overall squeezing effects.\\
Furthermore, we analyzed the behavior of particle correlation using the second-order correlation function and mode correlation using the JSI. Interestingly, our results reveal an inverse behavior between the second-order correlation and the JSI and consequently between particle and mode correlation. Additionally, also the entanglement between the signal and idler modes was investigated using Simons criterion and it has been shown that the mode entanglement rises with the pump power. However, since the multi-photon detection error $\varepsilon$ rises also with the pump power, the particle entanglement can only be utilized at a low pump power or a large $\Delta_{\mu, \mathrm{eff}}$. This indicates that either the particle or mode entanglement of a given mode can be predominantly utilized, a characteristic determined by the design and operation of the ring resonator. Nevertheless, for the entire quantum frequency comb, it is possible to generate modes exhibiting high mode entanglement as well as modes consisting of a high particle entanglement through clever dispersion engineering and thus precise control over the $\Delta_{\mu, \mathrm{eff}}$ of the individual modes.

\section{Summary}
\label{sec:summary-and-outlook}

In this work we derived closed-form analytical expressions describing the squeezing, the second-order correlation function, the JSI and the entanglement for each mode in a quantum frequency comb generated using a microring resonator. These equations enable the design and operation of microring resonators to enhance specific quantum features while suppressing others. We demonstrated that microring resonators can be used to generate a broad variety of quantum features for individual optical modes. Thus, the derived equations provide a foundational framework for various chip-integrated quantum optics applications, facilitating the development of quantum sensing, computing and communication.

\begin{acknowledgments}
	This work was supported by the IPCEI ME/CT project, which is supported by the Federal Ministry for Economic Affairs and Climate Action on the basis of a decision by the German Parliament, by the Ministry for Economic Affairs, Labor and Tourism of Baden-W\"urttemberg based on a decision of the State Parliament of Baden-W\"urttemberg, the Free State of Saxony on the basis of the budget adopted by the Saxon State Parliament, the Bavarian State Ministry for Economic Affairs, Regional Development and Energy and financed by the European Union - NextGenerationEU.
\end{acknowledgments}

\appendix

\section{Conversion of resonator modes to optical powers}\label{sec:conversion}

To convert the classical amplitude $\alpha$ of a resonator mode $a$ to optical powers, it is crucial to understand that $\alpha$ and $a$ describe the mode over the whole resonator volume. Thus, they are unitless and do not directly correspond to an optical power as the waveguide mode $b$, which corresponds to a flux in units of $\sqrt{\mathrm{Hz}}$. To solve this, the resonator modes are converted using the transmission rate $t$, which describes the amount of $\alpha$ that stays in the resonator using
\begin{equation}
	\alpha_{\sqrt{\mathrm{Hz}}}= \sqrt{t}\cdot\alpha, %
\end{equation}
which is the same as described in Appendix B of \cite{tritschler2024optical}. Then the amplitude can be converted to an optical power with the known relation
\begin{equation}\label{equ:power_transfer}
	\alpha_{\sqrt{\mathrm{Hz}}}= \sqrt{\frac{P}{\hbar\omega}}. 
\end{equation}
Note that the waveguide modes $b$ can always be transferred to an optical power using equation \ref{equ:power_transfer}.

\section{Input-output theory}\label{sec:input_output_theory}

To describe the outputs modes of the ringresonator $b_\mathrm{out}$ with equation \ref{equ:outputMode}, the input-output theory is performed. First, the equations of motion describing the signal and idler modes in the ringresonator are derived using the Hamiltonian of equation \ref{equ:fourWaveMixing_H} with
\begin{eqnarray}\label{equ:motion_in}
	\begin{aligned}
		\frac{d a_{s/i}(t)}{dt} =& - i \omega_{s/i} a_{s/i}(t) + \frac{\sigma}{2} a_{i/s}^{\dagger}(t) - \frac{\gamma}{2} a_{s/i}(t) - \frac{\kappa}{2} a_{s/i}(t)&  \\
		& + \sqrt{\kappa} b_{in, s/i}(t) + \sqrt{\gamma} b_{\gamma, s/i}(t).
	\end{aligned}
\end{eqnarray}
The boundary condition that has already been derived in \cite{Collet84, Gardiner85} is used to describe the output modes $b_{out}$ with
\begin{equation} \label{equ:boundary_condition}
	b_{in, s/i}(t) = \sqrt{\kappa} a_{s/i}(t) - b_{out, s/i}(t)
\end{equation}
and leads with equation \ref{equ:motion_in} to the equation of motion including the output modes with
\begin{eqnarray}\label{equ:motion_out}
\begin{aligned}
	\frac{d a_{s/i}(t)}{dt}  = &- i \omega_{s/i} a_{s/i}(t) + \frac{\sigma}{2} a_{i/s}^{\dagger}(t) - \frac{\gamma}{2} a_{s/i}(t) a_{s/i}(t)&\\
	& + \frac{\kappa}{2} - \sqrt{\kappa} b_{out, s/i}(t) + \sqrt{\gamma} b_{\gamma,s/i}(t).
\end{aligned}
\end{eqnarray}
For a more compact form, equations \ref{equ:motion_in} and \ref{equ:motion_out} are given in matrix notation with
\begin{eqnarray}\label{equ:matrix_in}
	\begin{aligned}
	\frac{d}{dt} \mathbf{A}(t) = (\mathbf{K}+\frac{\kappa}{2} \mathbbm{1}) \mathbf{A}(t) +  \sqrt{\kappa} \mathbf{B_{in}}(t) + \sqrt{\gamma} \mathbf{B_{\gamma}}(t), \label{equ:eqm_t_in}
\end{aligned}
\end{eqnarray}
and
\begin{eqnarray}\label{equ:matrix_out}
	\begin{aligned}
	\frac{d}{dt} \mathbf{A}(t) = (\mathbf{K}-\frac{\kappa}{2} \mathbbm{1}) \mathbf{A}(t) - \sqrt{\kappa} \mathbf{B_{out}}(t) + \sqrt{\gamma} \mathbf{B_{\gamma}}(t).\label{equ:eqm_t_out}
	\end{aligned}
\end{eqnarray}
Using the notation from equation \ref{equ:matrix_def} and \ref{equ:k_matrix} and the vector including the resonator modes
\begin{eqnarray}
\mathbf{A} = \left(\begin{array}{c}
	a_s\\
	a_s^{\dagger} \\
	a_i \\
	a_i^{\dagger}
\end{array}\right), \quad
\end{eqnarray} 
 the Fourier transformation of the equations \ref{equ:matrix_in} and \ref{equ:matrix_out} is performed, which leads to
\begin{eqnarray}\label{equ:matrix_out_ft}
	\begin{aligned}
	\mathbf{} [\mathbf{\Omega}-\mathbf{K}-\frac{\kappa}{2} \mathbf{I_4}] \mathbf{A(\omega)} = - \sqrt{\kappa} \mathbf{B_{out}}(\omega) + \sqrt{\gamma} \mathbf{B_{\gamma}}(\omega),
	\end{aligned}
\end{eqnarray}
\begin{eqnarray}\label{equ:matrix_in_ft}
	\begin{aligned}
		[\mathbf{\Omega}-\mathbf{K}+\frac{\kappa}{2} \mathbf{I_4}] \mathbf{A(\omega)} = \sqrt{\kappa} \mathbf{B_{in}}(\omega) + \sqrt{\gamma} \mathbf{B_{\gamma}}(\omega).
	\end{aligned}
\end{eqnarray}
Finally, the the resonator modes $\mathbf{A}$ are eliminated by combining equation \ref{equ:matrix_in_ft} and \ref{equ:matrix_out_ft}, which leads to equation \ref{equ:outputMode} of the main text.

\bibliography{QuantumFrequencyComb_bib}

@Article{micro_ringresonators,
  author   = {Bogaerts, W. and De Heyn, P. and Van Vaerenbergh, T. and De Vos, K. and Kumar Selvaraja, S. and Claes, T. and Dumon, P. and Bienstman, P. and Van Thourhout, D. and Baets, R.},
  journal  = {Laser \& Photonics Reviews},
  title    = {Silicon microring resonators},
  year     = {2012},
  number   = {1},
  pages    = {47-73},
  volume   = {6},
  doi      = {https://doi.org/10.1002/lpor.201100017},
}

@Article{Arrazola2021,
  author   = {Arrazola, J. M. and Bergholm, V. and Br{\'a}dler, K. and Bromley, T. R. and Collins, M. J. and Dhand, I. and Fumagalli, A. and Gerrits, T. and Goussev, A. and Helt, L. G. and Hundal, J. and Isacsson, T. and Israel, R. B. and Izaac, J. and Jahangiri, S. and Janik, R. and Killoran, N. and Kumar, S. P. and Lavoie, J. and Lita, A. E. and Mahler, D. H. and Menotti, M. and Morrison, B. and Nam, S. W. and Neuhaus, L. and Qi, H. Y. and Quesada, N. and Repingon, A. and Sabapathy, K. K. and Schuld, M. and Su, D. and Swinarton, J. and Sz{\'a}va, A. and Tan, K. and Tan, P. and Vaidya, V. D. and Vernon, Z. and Zabaneh, Z. and Zhang, Y.},
  journal  = {Nature},
  title    = {Quantum circuits with many photons on a programmable nanophotonic chip},
  year     = {2021},
  issn     = {1476-4687},
  month    = {Mar},
  number   = {7848},
  pages    = {54-60},
  volume   = {591},
  day      = {01},
  doi      = {10.1038/s41586-021-03202-1},
  url      = {https://doi.org/10.1038/s41586-021-03202-1},
}

@Article{Collet84,
  author    = {Collett, M. J. and Gardiner, C. W.},
  journal   = {Phys. Rev. A},
  title     = {Squeezing of intracavity and traveling-wave light fields produced in parametric amplification},
  year      = {1984},
  month     = {Sep},
  pages     = {1386--1391},
  volume    = {30},
  doi       = {10.1103/PhysRevA.30.1386},
  issue     = {3},
  numpages  = {0},
  publisher = {American Physical Society},
  url       = {https://link.aps.org/doi/10.1103/PhysRevA.30.1386},
}

@Article{Gardiner85,
  author    = {Gardiner, C. W. and Collett, M. J.},
  journal   = {Phys. Rev. A},
  title     = {Input and output in damped quantum systems: Quantum stochastic differential equations and the master equation},
  year      = {1985},
  month     = {Jun},
  pages     = {3761--3774},
  volume    = {31},
  doi       = {10.1103/PhysRevA.31.3761},
  issue     = {6},
  numpages  = {0},
  publisher = {American Physical Society},
  url       = {https://link.aps.org/doi/10.1103/PhysRevA.31.3761},
}

@Misc{zoller1997quantum,
  author        = {Peter Zoller and C. W. Gardiner},
  title         = {Quantum Noise in Quantum Optics: the Stochastic Schr\"odinger Equation},
  year          = {1997},
  archiveprefix = {arXiv},
  eprint        = {quant-ph/9702030},
  primaryclass  = {quant-ph},
}

@Article{PhysRevA.92.033840,
  author    = {Vernon, Z. and Sipe, J. E.},
  journal   = {Phys. Rev. A},
  title     = {Strongly driven nonlinear quantum optics in microring resonators},
  year      = {2015},
  month     = {Sep},
  pages     = {033840},
  volume    = {92},
  doi       = {10.1103/PhysRevA.92.033840},
  issue     = {3},
  numpages  = {17},
  publisher = {American Physical Society},
  url       = {https://link.aps.org/doi/10.1103/PhysRevA.92.033840},
}

@Article{FujiiTanabe+2020+1087+1104,
  author      = {Shun Fujii and Takasumi Tanabe},
  journal     = {Nanophotonics},
  title       = {Dispersion engineering and measurement of whispering gallery mode microresonator for {K}err frequency comb generation},
  year        = {2020},
  number      = {5},
  pages       = {1087--1104},
  volume      = {9},
  doi         = {doi:10.1515/nanoph-2019-0497},
  lastchecked = {2024-02-12},
  url         = {https://doi.org/10.1515/nanoph-2019-0497},
}

@Article{Chang2020,
  author   = {Chang, Lin and Xie, Weiqiang and Shu, Haowen and Yang, Qi-Fan and Shen, Boqiang and Boes, Andreas and Peters, Jon D. and Jin, Warren and Xiang, Chao and Liu, Songtao and Moille, Gregory and Yu, Su-Peng and Wang, Xingjun and Srinivasan, Kartik and Papp, Scott B. and Vahala, Kerry and Bowers, John E.},
  journal  = {Nature Communications},
  title    = {Ultra-efficient frequency comb generation in {AlGaAs}-on-insulator microresonators},
  year     = {2020},
  issn     = {2041-1723},
  month    = {Mar},
  number   = {1},
  pages    = {1331},
  volume   = {11},
  day      = {12},
  doi      = {10.1038/s41467-020-15005-5},
  url      = {https://doi.org/10.1038/s41467-020-15005-5},
}

@Article{tritschler2024optical,
  author    = {Tritschler, Patrick and Ohms, Torsten and Zimmermann, Andr\'e and Zschocke, Fabian and Strohm, Thomas and Degenfeld-Schonburg, Peter},
  journal   = {Phys. Rev. A},
  title     = {Optical interferometer using two-mode squeezed light for enhanced chip-integrated quantum metrology},
  year      = {2024},
  month     = {Jul},
  pages     = {012621},
  volume    = {110},
  doi       = {10.1103/PhysRevA.110.012621},
  issue     = {1},
  numpages  = {14},
  publisher = {American Physical Society},
  url       = {https://link.aps.org/doi/10.1103/PhysRevA.110.012621},
}

@Article{Kippenberg2011-uh,
  author    = {Kippenberg, T J and Holzwarth, R and Diddams, S A},
  journal   = {Science},
  title     = {Microresonator-based optical frequency combs},
  year      = {2011},
  month     = apr,
  number    = {6029},
  pages     = {555--559},
  volume    = {332},
  publisher = {American Association for the Advancement of Science (AAAS)},
}

@Article{Del’Haye2007,
  author   = {Del'Haye, P. and Schliesser, A. and Arcizet, O. and Wilken, T. and Holzwarth, R. and Kippenberg, T. J.},
  journal  = {Nature},
  title    = {Optical frequency comb generation from a monolithic microresonator},
  year     = {2007},
  issn     = {1476-4687},
  month    = {Dec},
  number   = {7173},
  pages    = {1214-1217},
  volume   = {450},
  day      = {01},
  doi      = {10.1038/nature06401},
  url      = {https://doi.org/10.1038/nature06401},
}

@Article{Fortier2019,
  author   = {Fortier, Tara and Baumann, Esther},
  journal  = {Communications Physics},
  title    = {20 years of developments in optical frequency comb technology and applications},
  year     = {2019},
  issn     = {2399-3650},
  month    = {Dec},
  number   = {1},
  pages    = {153},
  volume   = {2},
  day      = {06},
  doi      = {10.1038/s42005-019-0249-y},
  url      = {https://doi.org/10.1038/s42005-019-0249-y},
}

@Article{Liu:16,
  author    = {Nannan Liu and Yuhong Liu and Jiamin Li and Lei Yang and Xiaoying Li},
  journal   = {Opt. Express},
  title     = {Generation of multi-mode squeezed vacuum using pulse pumped fiber optical parametric amplifiers},
  year      = {2016},
  month     = {Feb},
  number    = {3},
  pages     = {2125--2133},
  volume    = {24},
  doi       = {10.1364/OE.24.002125},
  publisher = {Optica Publishing Group},
  url       = {https://opg.optica.org/oe/abstract.cfm?URI=oe-24-3-2125},
}

@Article{PhysRevLett.114.050501,
  author    = {Gerke, S. and Sperling, J. and Vogel, W. and Cai, Y. and Roslund, J. and Treps, N. and Fabre, C.},
  journal   = {Phys. Rev. Lett.},
  title     = {Full Multipartite Entanglement of Frequency-Comb Gaussian States},
  year      = {2015},
  month     = {Feb},
  pages     = {050501},
  volume    = {114},
  doi       = {10.1103/PhysRevLett.114.050501},
  issue     = {5},
  numpages  = {5},
  publisher = {American Physical Society},
  url       = {https://link.aps.org/doi/10.1103/PhysRevLett.114.050501},
}

@Article{Guidry2022,
  author   = {Guidry, Melissa A. and Lukin, Daniil M. and Yang, Ki Youl and Trivedi, Rahul and Vu{\v{c}}kovi{\'{c}}, Jelena},
  journal  = {Nature Photonics},
  title    = {Quantum optics of soliton microcombs},
  year     = {2022},
  issn     = {1749-4893},
  month    = {Jan},
  number   = {1},
  pages    = {52-58},
  volume   = {16},
  day      = {01},
  doi      = {10.1038/s41566-021-00901-z},
  url      = {https://doi.org/10.1038/s41566-021-00901-z},
}

@Article{PhysRevLett.101.130501,
  author    = {Menicucci, Nicolas C. and Flammia, Steven T. and Pfister, Olivier},
  journal   = {Phys. Rev. Lett.},
  title     = {One-Way Quantum Computing in the Optical Frequency Comb},
  year      = {2008},
  month     = {Sep},
  pages     = {130501},
  volume    = {101},
  doi       = {10.1103/PhysRevLett.101.130501},
  issue     = {13},
  numpages  = {4},
  publisher = {American Physical Society},
  url       = {https://link.aps.org/doi/10.1103/PhysRevLett.101.130501},
}

@Article{Strekalov_2016,
  author    = {Dmitry V Strekalov and Christoph Marquardt and Andrey B Matsko and Harald G L Schwefel and Gerd Leuchs},
  journal   = {Journal of Optics},
  title     = {Nonlinear and quantum optics with whispering gallery resonators},
  year      = {2016},
  month     = {nov},
  number    = {12},
  pages     = {123002},
  volume    = {18},
  doi       = {10.1088/2040-8978/18/12/123002},
  publisher = {IOP Publishing},
  url       = {https://dx.doi.org/10.1088/2040-8978/18/12/123002},
}

@Article{Lu2019,
  author   = {Lu, Xiyuan and Li, Qing and Westly, Daron A. and Moille, Gregory and Singh, Anshuman and Anant, Vikas and Srinivasan, Kartik},
  journal  = {Nature Physics},
  title    = {Chip-integrated visible--telecom entangled photon pair source for quantum communication},
  year     = {2019},
  issn     = {1745-2481},
  month    = {Apr},
  number   = {4},
  pages    = {373-381},
  volume   = {15},
  day      = {01},
  doi      = {10.1038/s41567-018-0394-3},
  url      = {https://doi.org/10.1038/s41567-018-0394-3},
}

@Article{Yang2021,
  author   = {Yang, Zijiao and Jahanbozorgi, Mandana and Jeong, Dongin and Sun, Shuman and Pfister, Olivier and Lee, Hansuek and Yi, Xu},
  journal  = {Nature Communications},
  title    = {A squeezed quantum microcomb on a chip},
  year     = {2021},
  issn     = {2041-1723},
  month    = {Aug},
  number   = {1},
  pages    = {4781},
  volume   = {12},
  day      = {06},
  doi      = {10.1038/s41467-021-25054-z},
  url      = {https://doi.org/10.1038/s41467-021-25054-z},
}

@Article{PhysRevLett.108.083601,
  author    = {Pinel, Olivier and Jian, Pu and de Ara\'ujo, Renn\'e Medeiros and Feng, Jinxia and Chalopin, Beno\^{\i}t and Fabre, Claude and Treps, Nicolas},
  journal   = {Phys. Rev. Lett.},
  title     = {Generation and Characterization of Multimode Quantum Frequency Combs},
  year      = {2012},
  month     = {Feb},
  pages     = {083601},
  volume    = {108},
  doi       = {10.1103/PhysRevLett.108.083601},
  issue     = {8},
  numpages  = {5},
  publisher = {American Physical Society},
  url       = {https://link.aps.org/doi/10.1103/PhysRevLett.108.083601},
}

@Article{Papp:14,
  author    = {Scott B. Papp and Katja Beha and Pascal Del'Haye and Franklyn Quinlan and Hansuek Lee and Kerry J. Vahala and Scott A. Diddams},
  journal   = {Optica},
  title     = {Microresonator frequency comb optical clock},
  year      = {2014},
  month     = {Jul},
  number    = {1},
  pages     = {10--14},
  volume    = {1},
  doi       = {10.1364/OPTICA.1.000010},
  publisher = {Optica Publishing Group},
  url       = {https://opg.optica.org/optica/abstract.cfm?URI=optica-1-1-10},
}

@Article{Fortier_2024,
  author    = {Fortier, Tara and as a representative of the BACON collaboration},
  journal   = {Journal of Physics: Conference Series},
  title     = {Frequency combs for precision synthesis and characterization of optical atomic standards},
  year      = {2024},
  month     = {nov},
  number    = {1},
  pages     = {012021},
  volume    = {2889},
  doi       = {10.1088/1742-6596/2889/1/012021},
  publisher = {IOP Publishing},
  url       = {https://dx.doi.org/10.1088/1742-6596/2889/1/012021},
}

@Article{Zhang2015,
  author   = {Zhang, S. Y. and Wu, J. T. and Zhang, Y. L. and Leng, J. X. and Yang, W. P. and Zhang, Z. G. and Zhao, J. Y.},
  journal  = {Scientific Reports},
  title    = {Direct frequency comb optical frequency standard based on two-photon transitions of thermal atoms},
  year     = {2015},
  issn     = {2045-2322},
  month    = {Oct},
  number   = {1},
  pages    = {15114},
  volume   = {5},
  day      = {13},
  doi      = {10.1038/srep15114},
  url      = {https://doi.org/10.1038/srep15114},
}

@Article{Picque2019,
  author   = {Picqu{\'e}, Nathalie and H{\"a}nsch, Theodor W.},
  journal  = {Nature Photonics},
  title    = {Frequency comb spectroscopy},
  year     = {2019},
  issn     = {1749-4893},
  month    = {Mar},
  number   = {3},
  pages    = {146-157},
  volume   = {13},
  day      = {01},
  doi      = {10.1038/s41566-018-0347-5},
  url      = {https://doi.org/10.1038/s41566-018-0347-5},
}

@Article{Kubo1962,
  author    = {Kubo , Ryogo},
  journal   = {Journal of the Physical Society of Japan},
  title     = {Generalized Cumulant Expansion Method},
  year      = {1962},
  issn      = {0031-9015},
  month     = {Jul},
  number    = {7},
  pages     = {1100-1120},
  volume    = {17},
  day       = {15},
  doi       = {10.1143/JPSJ.17.1100},
  publisher = {The Physical Society of Japan},
  url       = {https://doi.org/10.1143/JPSJ.17.1100},
}

@Article{PhysRevLett.107.233002,
  author    = {Foltynowicz, Aleksandra and Ban, Ticijana and Mas\l{}owski, Piotr and Adler, Florian and Ye, Jun},
  journal   = {Phys. Rev. Lett.},
  title     = {Quantum-Noise-Limited Optical Frequency Comb Spectroscopy},
  year      = {2011},
  month     = {Nov},
  pages     = {233002},
  volume    = {107},
  doi       = {10.1103/PhysRevLett.107.233002},
  issue     = {23},
  numpages  = {5},
  publisher = {American Physical Society},
  url       = {https://link.aps.org/doi/10.1103/PhysRevLett.107.233002},
}

@Article{Kuse2019,
  author   = {Kuse, N. and Fermann, M. E.},
  journal  = {APL Photonics},
  title    = {Frequency-modulated comb {LIDAR}},
  year     = {2019},
  issn     = {2378-0967},
  month    = {Oct},
  number   = {10},
  pages    = {106105},
  volume   = {4},
  day      = {09},
  doi      = {10.1063/1.5120321},
  url      = {https://doi.org/10.1063/1.5120321},
}

@Article{Horiuchi2024,
  author  = {Horiuchi, Noriaki},
  journal = {Nature Photonics},
  title   = {Optical frequency comb for multi-sensors},
  year    = {2024},
  issn    = {1749-4893},
  month   = {Jul},
  number  = {7},
  pages   = {648-648},
  volume  = {18},
  day     = {01},
  doi     = {10.1038/s41566-024-01464-5},
  url     = {https://doi.org/10.1038/s41566-024-01464-5},
}

@Article{Herr2012,
  author   = {Herr, T. and Hartinger, K. and Riemensberger, J. and Wang, C. Y. and Gavartin, E. and Holzwarth, R. and Gorodetsky, M. L. and Kippenberg, T. J.},
  journal  = {Nature Photonics},
  title    = {Universal formation dynamics and noise of {K}err-frequency combs in microresonators},
  year     = {2012},
  issn     = {1749-4893},
  month    = {Jul},
  number   = {7},
  pages    = {480-487},
  volume   = {6},
  day      = {01},
  doi      = {10.1038/nphoton.2012.127},
  url      = {https://doi.org/10.1038/nphoton.2012.127},
}

@Article{Cheng2023,
  author   = {Cheng, Xiang and Chang, Kai-Chi and Sarihan, Murat Can and Mueller, Andrew and Spiropulu, Maria and Shaw, Matthew D. and Korzh, Boris and Faraon, Andrei and Wong, Franco N. C. and Shapiro, Jeffrey H. and Wong, Chee Wei},
  journal  = {Communications Physics},
  title    = {High-dimensional time-frequency entanglement in a singly-filtered biphoton frequency comb},
  year     = {2023},
  issn     = {2399-3650},
  month    = {Sep},
  number   = {1},
  pages    = {278},
  volume   = {6},
  day      = {28},
  doi      = {10.1038/s42005-023-01370-2},
  url      = {https://doi.org/10.1038/s42005-023-01370-2},
}

@Article{Caspani2017,
  author   = {Caspani, Lucia and Xiong, Chunle and Eggleton, Benjamin J. and Bajoni, Daniele and Liscidini, Marco and Galli, Matteo and Morandotti, Roberto and Moss, David J.},
  journal  = {Light: Science {\&} Applications},
  title    = {Integrated sources of photon quantum states based on nonlinear optics},
  year     = {2017},
  issn     = {2047-7538},
  month    = {Nov},
  number   = {11},
  pages    = {e17100-e17100},
  volume   = {6},
  day      = {01},
  doi      = {10.1038/lsa.2017.100},
  url      = {https://doi.org/10.1038/lsa.2017.100},
}

@Article{PhysRevA.93.053802,
  author    = {Lucivero, Vito Giovanni and Jim\'enez-Mart\'{\i}nez, Ricardo and Kong, Jia and Mitchell, Morgan W.},
  journal   = {Phys. Rev. A},
  title     = {Squeezed-light spin noise spectroscopy},
  year      = {2016},
  month     = {May},
  pages     = {053802},
  volume    = {93},
  doi       = {10.1103/PhysRevA.93.053802},
  issue     = {5},
  numpages  = {7},
  publisher = {American Physical Society},
  url       = {https://link.aps.org/doi/10.1103/PhysRevA.93.053802},
}

@Article{2025Herman,
  author    = {Herman, Daniel I. and Walsh, Mathieu and Kreider, Molly Kate and Lordi, Noah and Tsao, Eugene J. and Lind, Alexander J. and Heyrich, Matthew and Combes, Joshua and Genest, J{\'e}r{\^o}me and Diddams, Scott A.},
  journal   = {Science},
  title     = {Squeezed dual-comb spectroscopy},
  year      = {2025},
  month     = {2025/02/02},
  number    = {0},
  pages     = {eads6292},
  volume    = {0},
  doi       = {10.1126/science.ads6292},
  publisher = {American Association for the Advancement of Science},
  url       = {https://doi.org/10.1126/science.ads6292},
}

@Article{Michael2019,
  author   = {Michael, Yoad and Bello, Leon and Rosenbluh, Michael and Pe'er, Avi},
  journal  = {npj Quantum Information},
  title    = {Squeezing-enhanced {R}aman spectroscopy},
  year     = {2019},
  issn     = {2056-6387},
  month    = {Oct},
  number   = {1},
  pages    = {81},
  volume   = {5},
  day      = {01},
  doi      = {10.1038/s41534-019-0197-0},
  url      = {https://doi.org/10.1038/s41534-019-0197-0},
}

@InBook{Walls2008,
  author    = {Walls, D.F. and Milburn, Gerard J.},
  editor    = {Walls, D.F. and Milburn, Gerard J.},
  pages     = {7--27},
  publisher = {Springer Berlin Heidelberg},
  title     = {Quantisation of the Electromagnetic Field},
  year      = {2008},
  address   = {Berlin, Heidelberg},
  isbn      = {978-3-540-28574-8},
  booktitle = {Quantum Optics},
  doi       = {10.1007/978-3-540-28574-8_2},
  url       = {https://doi.org/10.1007/978-3-540-28574-8_2},
}

@Misc{tritschler2025chipintegratedsqueezedlight,
  author        = {Patrick Tritschler and Torsten Ohms and Christian Schweikert and Onur Sözen and Rouven H. Klenk and Simon Abdani and Wolfgang Vogel and Georg Rademacher and André Zimmermann and Peter Degenfeld-Schonburg},
  title         = {Chip-integrated single-mode coherent-squeezed light source using four-wave mixing in microresonators},
  year          = {2025},
  archiveprefix = {arXiv},
  eprint        = {2502.16278},
  primaryclass  = {quant-ph},
  url           = {https://arxiv.org/abs/2502.16278},
}

@Article{Bourassa_2021,
  author    = {Bourassa, J. Eli and Alexander, Rafael N. and Vasmer, Michael and Patil, Ashlesha and Tzitrin, Ilan and Matsuura, Takaya and Su, Daiqin and Baragiola, Ben Q. and Guha, Saikat and Dauphinais, Guillaume and Sabapathy, Krishna K. and Menicucci, Nicolas C. and Dhand, Ish},
  journal   = {Quantum},
  title     = {Blueprint for a Scalable Photonic Fault-Tolerant Quantum Computer},
  year      = {2021},
  issn      = {2521-327X},
  month     = feb,
  pages     = {392},
  volume    = {5},
  doi       = {10.22331/q-2021-02-04-392},
  publisher = {Verein zur Forderung des Open Access Publizierens in den Quantenwissenschaften},
  url       = {http://dx.doi.org/10.22331/q-2021-02-04-392},
}

@Article{PhysRevA.107.052414,
  author    = {Fukui, Kosuke},
  journal   = {Phys. Rev. A},
  title     = {High-threshold fault-tolerant quantum computation with the Gottesman-Kitaev-Preskill qubit under noise in an optical setup},
  year      = {2023},
  month     = {May},
  pages     = {052414},
  volume    = {107},
  doi       = {10.1103/PhysRevA.107.052414},
  issue     = {5},
  numpages  = {11},
  publisher = {American Physical Society},
  url       = {https://link.aps.org/doi/10.1103/PhysRevA.107.052414},
}

@Article{Jia_2024,
  author    = {Jia, Wenxuan, et. al.},
  journal   = {Science},
  title     = {Squeezing the quantum noise of a gravitational-wave detector below the standard quantum limit},
  year      = {2024},
  issn      = {1095-9203},
  month     = sep,
  number    = {6715},
  pages     = {1318–1321},
  volume    = {385},
  doi       = {10.1126/science.ado8069},
  publisher = {American Association for the Advancement of Science (AAAS)},
  url       = {http://dx.doi.org/10.1126/science.ado8069},
}

@Article{Goda2008,
  author   = {Goda, K. and Miyakawa, O. and Mikhailov, E. E. and Saraf, S. and Adhikari, R. and McKenzie, K. and Ward, R. and Vass, S. and Weinstein, A. J. and Mavalvala, N.},
  journal  = {Nature Physics},
  title    = {A quantum-enhanced prototype gravitational-wave detector},
  year     = {2008},
  issn     = {1745-2481},
  month    = {Jun},
  number   = {6},
  pages    = {472-476},
  volume   = {4},
  day      = {01},
  doi      = {10.1038/nphys920},
  url      = {https://doi.org/10.1038/nphys920},
}

@Article{Vahlbruch_2010,
  author   = {Vahlbruch, Henning and Khalaidovski, Alexander and Lastzka, Nico and Gräf, Christian and Danzmann, Karsten and Schnabel, Roman},
  journal  = {Classical and Quantum Gravity},
  title    = {The {GEO 600} squeezed light source},
  year     = {2010},
  month    = {apr},
  number   = {8},
  pages    = {084027},
  volume   = {27},
  doi      = {10.1088/0264-9381/27/8/084027},
  url      = {https://dx.doi.org/10.1088/0264-9381/27/8/084027},
}

@Article{Lugiato_2018,
  author    = {Lugiato, L. A. and Prati, F. and Gorodetsky, M. L. and Kippenberg, T. J.},
  journal   = {Philosophical Transactions of the Royal Society A: Mathematical, Physical and Engineering Sciences},
  title     = {From the {L}ugiato–{L}efever equation to microresonator-based soliton {K}err frequency combs},
  year      = {2018},
  issn      = {1471-2962},
  month     = nov,
  number    = {2135},
  pages     = {20180113},
  volume    = {376},
  doi       = {10.1098/rsta.2018.0113},
  publisher = {The Royal Society},
  url       = {http://dx.doi.org/10.1098/rsta.2018.0113},
}

@Article{Coen:13,
  author    = {St\'{e}phane Coen and Hamish G. Randle and Thibaut Sylvestre and Miro Erkintalo},
  journal   = {Opt. Lett.},
  title     = {Modeling of octave-spanning {K}err frequency combs using a generalized mean-field {L}ugiato-{L}efever model},
  year      = {2013},
  month     = {Jan},
  number    = {1},
  pages     = {37--39},
  volume    = {38},
  doi       = {10.1364/OL.38.000037},
  publisher = {Optica Publishing Group},
  url       = {https://opg.optica.org/ol/abstract.cfm?URI=ol-38-1-37},
}

@Article{PhysRevLett.134.123802,
  author    = {Tritschler, Patrick and Schweikert, Christian and Klenk, Rouven H. and Abdani, Simon and S\"ozen, Onur and Vogel, Wolfgang and Rademacher, Georg and Ohms, Torsten and Zimmermann, Andr\'e and Degenfeld-Schonburg, Peter},
  journal   = {Phys. Rev. Lett.},
  title     = {Nonlinear Optical Bistability in Microring Resonators for Enhanced Phase Sensing},
  year      = {2025},
  month     = {Mar},
  pages     = {123802},
  volume    = {134},
  doi       = {10.1103/PhysRevLett.134.123802},
  issue     = {12},
  numpages  = {9},
  publisher = {American Physical Society},
  url       = {https://link.aps.org/doi/10.1103/PhysRevLett.134.123802},
}

@Book{wiseman_milburn_2009,
  author    = {Wiseman, Howard M. and Milburn, Gerard J.},
  publisher = {Cambridge University Press},
  title     = {Quantum Measurement and Control},
  year      = {2009},
  doi       = {10.1017/CBO9780511813948},
  place     = {Cambridge},
}

@Article{Zielnicki2018,
  author    = {Kevin Zielnicki and Karina Garay-Palmett and Daniel Cruz-Delgado and Hector Cruz-Ramirez and Michael F. O’Boyle and Bin Fang and Virginia O. Lorenz and Alfred B. U’Ren and Paul G. Kwiat},
  journal   = {Journal of Modern Optics},
  title     = {Joint spectral characterization of photon-pair sources},
  year      = {2018},
  number    = {10},
  pages     = {1141-1160},
  volume    = {65},
  doi       = {10.1080/09500340.2018.1437228},
  eprint    = {https://doi.org/10.1080/09500340.2018.1437228},
  publisher = {Taylor & Francis},
  url       = {https://doi.org/10.1080/09500340.2018.1437228},
}

@Article{Simon_2000,
  author    = {Simon, R.},
  journal   = {Physical Review Letters},
  title     = {Peres-Horodecki Separability Criterion for Continuous Variable Systems},
  year      = {2000},
  issn      = {1079-7114},
  month     = mar,
  number    = {12},
  pages     = {2726–2729},
  volume    = {84},
  doi       = {10.1103/physrevlett.84.2726},
  publisher = {American Physical Society (APS)},
  url       = {http://dx.doi.org/10.1103/PhysRevLett.84.2726},
}

@Article{Peres_1996,
  author    = {Peres, Asher},
  journal   = {Physical Review Letters},
  title     = {Separability Criterion for Density Matrices},
  year      = {1996},
  issn      = {1079-7114},
  month     = aug,
  number    = {8},
  pages     = {1413–1415},
  volume    = {77},
  doi       = {10.1103/physrevlett.77.1413},
  publisher = {American Physical Society (APS)},
  url       = {http://dx.doi.org/10.1103/PhysRevLett.77.1413},
}

@Article{Horodecki_1997,
  author    = {Horodecki, Pawel},
  journal   = {Physics Letters A},
  title     = {Separability criterion and inseparable mixed states with positive partial transposition},
  year      = {1997},
  issn      = {0375-9601},
  month     = aug,
  number    = {5},
  pages     = {333–339},
  volume    = {232},
  doi       = {10.1016/s0375-9601(97)00416-7},
  publisher = {Elsevier BV},
  url       = {http://dx.doi.org/10.1016/S0375-9601(97)00416-7},
}

@Article{Weedbrook_2012,
  author    = {Weedbrook, Christian and Pirandola, Stefano and García-Patrón, Raúl and Cerf, Nicolas J. and Ralph, Timothy C. and Shapiro, Jeffrey H. and Lloyd, Seth},
  journal   = {Reviews of Modern Physics},
  title     = {Gaussian quantum information},
  year      = {2012},
  issn      = {1539-0756},
  month     = may,
  number    = {2},
  pages     = {621–669},
  volume    = {84},
  doi       = {10.1103/revmodphys.84.621},
  publisher = {American Physical Society (APS)},
  url       = {http://dx.doi.org/10.1103/RevModPhys.84.621},
}

@Article{Serafini_2003,
  author    = {Serafini, Alessio and Illuminati, Fabrizio and Siena, Silvio De},
  journal   = {Journal of Physics B: Atomic, Molecular and Optical Physics},
  title     = {Symplectic invariants, entropic measures and correlations of Gaussian states},
  year      = {2003},
  issn      = {1361-6455},
  month     = dec,
  number    = {2},
  pages     = {L21–L28},
  volume    = {37},
  doi       = {10.1088/0953-4075/37/2/l02},
  publisher = {IOP Publishing},
  url       = {http://dx.doi.org/10.1088/0953-4075/37/2/L02},
}

@Article{Gisin_2002,
  author    = {Gisin, Nicolas and Ribordy, Grégoire and Tittel, Wolfgang and Zbinden, Hugo},
  journal   = {Reviews of Modern Physics},
  title     = {Quantum cryptography},
  year      = {2002},
  issn      = {1539-0756},
  month     = mar,
  number    = {1},
  pages     = {145–195},
  volume    = {74},
  doi       = {10.1103/revmodphys.74.145},
  publisher = {American Physical Society (APS)},
  url       = {http://dx.doi.org/10.1103/RevModPhys.74.145},
}

@Article{Reimer2016,
  author    = {Reimer, Christian and Kues, Michael and Roztocki, Piotr and Wetzel, Benjamin and Grazioso, Fabio and Little, Brent E. and Chu, Sai T. and Johnston, Tudor and Bromberg, Yaron and Caspani, Lucia and Moss, David J. and Morandotti, Roberto},
  journal   = {Science},
  title     = {Generation of multiphoton entangled quantum states by means of integrated frequency combs},
  year      = {2016},
  month     = {Mar},
  number    = {6278},
  pages     = {1176-1180},
  volume    = {351},
  day       = {11},
  doi       = {10.1126/science.aad8532},
  publisher = {American Association for the Advancement of Science},
  url       = {https://doi.org/10.1126/science.aad8532},
}

\end{document}